\documentclass[twocolumn,astrosym,twocolappendix]{aastex631}

\usepackage{apjfonts}

\usepackage{natbib}
\usepackage{afterpage}
\usepackage{tensor}
\usepackage{graphicx}
\usepackage{subfigure}
\usepackage{amsmath, amssymb, amsfonts, mathtools}
\usepackage{bm}

\DeclareMathSymbol{v}{\mathalpha}{cmletters}{"76}

\usepackage{ulem}
\usepackage{xcolor}
\usepackage{enumitem}
\usepackage{hyperref}

\newcommand \rg    {\ensuremath{ r_{\rm g}  }}
\newcommand \rNS   {\ensuremath{ r_* }}

\newcommand \rout  {\ensuremath{ r_{\rm domain} }}
\newcommand \rmag  {\ensuremath{ r_{\rm m} }}
\newcommand \rmax {\ensuremath{ r_{\rm max} }}
\newcommand \rinner {\ensuremath{ r_{\rm in} }}
\newcommand \rco  {\ensuremath{ r_{\rm co} }}
\newcommand \RLC   {\ensuremath{ R_{\rm LC} }}
\newcommand \rlc   {\ensuremath{ \RLC } }
\newcommand \MNS   {\ensuremath{ M }}
\newcommand \nphihi {$N_{\phi}^{\rm high}$}
\newcommand \rgu    {\ensuremath{ \, \rg  }}
\newcommand \rgc    {\ensuremath{ \, \rg/c  }}
\newcommand \crg    {\ensuremath{ \, c/\rg  }}
\newcommand \Omegastar   {\ensuremath{ \Omega_* }}
\newcommand \Pstar   {\ensuremath{ P_* }}
\newcommand \Omegak   {\ensuremath{ \Omega_{\rm K} }}
\newcommand \Omegarho   {\ensuremath{ \langle\Omega\rangle_\rho }}
\newcommand \Omegab   {\ensuremath{ \langle\Omega\rangle_{b^2} }}

\newcommand \rhobg  {\ensuremath{ \rho_{\rm bg} }}
\newcommand \rhom  {\ensuremath{ \rho_{\rm m} }}
\newcommand \rhoaf  {\ensuremath{ \rho_{\rm af} }}
\newcommand \ubg  {\ensuremath{ \epsilon_{\rm bg} }}
\newcommand \Dp  {\ensuremath{ \mathrm{d}p }}
\newcommand \us  {\ensuremath{ u_{\rm s} }}
\newcommand \bs  {\ensuremath{ b_{\rm s} }}

\newcommand \ff  {\ensuremath{ \psi  }}
\newcommand \ffo  {\ensuremath{ \ff_{\rm open, 0}  }}
\newcommand \ffopen {\ensuremath{ \ff_{\rm open} }}
\newcommand \fftot  {\ensuremath{ \ff_{\rm tot} }}
\newcommand \Lj  {\ensuremath{ L_{\rm jet}  }}
\newcommand \Lo  {\ensuremath{ L_0  }}
\newcommand \No  {\ensuremath{ N_0  }}
\newcommand \Nem  {\ensuremath{ N_{\rm EM}  }}
\newcommand \Nhydro  {\ensuremath{ N_{\rm hydro}  }}

\newcommand \mdot  {\ensuremath{ \dot{M}  }}
\newcommand \dtsteady  {\ensuremath{ \Delta t_{\rm steady}  }}

\renewcommand{\vec}[1]{\boldsymbol{#1}}

\begin{document}

\title{Accreting Neutron Stars in 3D GRMHD Simulations: Jets, Magnetic Polarity,\\ and the Interchange Slingshot}

\author[0000-0001-6173-0099]{Kyle Parfrey}
\affiliation{Princeton Plasma Physics Laboratory, Princeton, NJ 08540, USA}
\email{kparfrey@pppl.gov}

\author[0000-0002-9182-2047]{Alexander Tchekhovskoy}
\affiliation{Center for Interdisciplinary Exploration \& Research in Astrophysics (CIERA), Physics \& Astronomy, Northwestern University, Evanston, IL 60201, USA}

\begin{abstract}

%176 of 250 words
Accreting neutron stars differ from black holes by the presence of the star's own magnetic field, whose interaction with the accretion flow is a central component in understanding these systems' disk structure, outflows, jets, and spin evolution. It also introduces an additional degree of freedom, as the stellar dipole can have any orientation relative to the inner disk's magnetic field. We present a suite of 3D general-relativistic magnetohydrodynamic (GRMHD) simulations in which we investigate the two extreme polarities, with the dipole field being either parallel or antiparallel to the initial disk field, in both the accreting and propeller states. When the magnetosphere truncates the disk near or beyond the corotation radius, most of the system's properties, including the relativistic jet power, are independent of the star--disk relative polarity. However, when the disk extends well inside the corotation radius, in the parallel orientation the jet power is suppressed and the inner disk is less dense and more strongly magnetized. We suggest a physical mechanism that may account for this behavior --- the interchange slingshot --- and discuss its astrophysical implications.

\end{abstract}


\section{Introduction}

The interaction between the magnetic field frozen into a neutron star's solid crust and a surrounding accretion flow leads to a range of observed behaviors distinct from their black-hole cousins, among them X-ray pulsations from streams of infalling plasma confined by the star's field \citep{1971ApJ...167L..67G,1972A&A....21....1P,1976MNRAS.175..395B,1989PASJ...41....1N}, centrifugal inhibition of accretion by the rotating magnetosphere \citep[the ``propeller effect'';][]{1975A&A....39..185I,1986ApJ...308..669S,2001ApJ...561..924C,2016A&A...593A..16T}, and transitions between accretion-powered X-ray pulsar and rotation-powered radio pulsar states \citep{2009Sci...324.1411A,2013Natur.501..517P,2015ApJ...806..148B}. Pulsations and burst oscillations \citep{2000ARA&A..38..717V,2012ARA&A..50..609W} give spin measurements for many systems that are much more precise and robust than are available for any black hole, and sometimes even indicate reversals in the sign of the stellar torque \citep{1988Natur.333..746M,1989ApJ...336..376D,1997ApJ...474..414C}. 

Like black holes, neutron stars launch relativistic jets, which are occasionally resolvable \citep{2001ApJ...558..283F,2004Natur.427..222F} but are generally inferred from the presence of continuum radio emission \citep{2006MNRAS.366...79M,2017MNRAS.470..324T,2021MNRAS.507.3899V}. For both accretor classes the jets may be launched magnetocentrifugally by the accretion disk \citep{1982MNRAS.199..883B} or by a rotating central compact object threaded by a magnetic field. For neutron stars this latter channel relies on the collimation of the \citet{1969ApJ...157..869G} electromagnetic pulsar wind by the accretion flow and its associated outflows. The detection of jet-like radio emission from a strongly magnetized neutron star \citep{2018Natur.562..233V} weighs against a disk-powered mechanism, as the star's magnetic field is expected to truncate the accretion flow far from the region of relativistic orbital velocities \citep{2008A&A...477....1M}.

Theoretical models have been proposed in which the star and disk remain coupled by the star's magnetic field to large distances \citep{1977ApJ...215..897E,1978ApJ...223L..83G,1987A&A...183..257W,1995ApJ...449L.153W} or when the star--disk connection is limited to a region near the inner edge of the disk \citep{1994ApJ...429..781S,1995MNRAS.275..244L,2005ApJ...632L.135M}.
A basic model of how the strength of a neutron star's Goldreich--Julian electromagnetic wind, and hence the relativistic jet power, can be increased by stellar flux opening due to interaction with an accretion flow was suggested by \citet{2016ApJ...822...33P}.

Accretion onto magnetized stars has been studied extensively in the non-relativistic regime since the pioneering work of \citet{1996ApJ...468L..37H} and \citet{1997ApJ...489..890M}. Axisymmetric simulations have studied funnel flows \citep{2002ApJ...578..420R,2008A&A...478..155B} and the propeller regime \citep{2004ApJ...616L.151R,2006ApJ...646..304U}; focused on extensive star--disk coupling \citep{2009A&A...508.1117Z} or ejections driven by inflating field lines \citep{2013A&A...550A..99Z}; and created jets with magnetic towers \citep{2004ApJ...605..307K}. The first 3D simulations used resistive $\alpha$--disk prescriptions to study accretion onto stars with spin--magnetic misalignment \citep{2003ApJ...595.1009R} before attention largely switched to a self-consistent ideal-MHD approach \citep{2012MNRAS.421...63R, 2023arXiv230915318Z}.

In the relativistic regime, idealized simulations coupling a force-free magnetosphere to a prescribed disk \citep{2017MNRAS.469.3656P} preceded full GRMHD studies of accretion onto rotating stars \citep[henceforth PT17]{2017ApJ...851L..34P}. \citet{2022MNRAS.515.3144D} investigated accretion onto rotating stars with multipolar magnetic fields, while radiation-GRMHD simulations have been employed to model super-Eddington accretion onto non-rotating stars, with application to the ULX pulsars \citep{2017ApJ...845L...9T,2021ApJ...917L..31A, 2022MNRAS.517.3212C,2023ApJ...952...62I}. In this paper we present the first relativistic 3D simulations of rotating, magnetized stars interacting with accretion flows.

\section{Numerical Approach and Problem Configuration}
\label{sec:setup}

\subsection{Physical quantities}

We use a variant of the \textsc{harmpi} finite-volume GRMHD code, based on the original \textsc{harm} of \citet{2003ApJ...589..444G} and \citet{2006ApJ...641..626N}, that has been modified for increased stability in strongly magnetized regions. This method, which in effect combines MHD and force-free regions in a single self-consistent simulation, is outlined briefly in \hyperlink{PT17target}{PT17} and described in greater depth in Appendix~\ref{sec:app_method}. 
We evolve the ideal GRMHD equations,
\begin{subequations}
\begin{align}
  \nabla_\mu (\rho u^\mu)&=0,\\
  \nabla_\mu \tensor{T}{^{\mu}_{\nu}} &=0,\\
  \nabla_\mu \tensor[^*]{F}{^{\mu\nu}}  &=0,
\end{align}
\label{eq:mhd}
\end{subequations}
where 
\begin{multline}
\label{eq:tmunu}
  %%% Gaussian units
  %%%T^{\mu\nu}=\left(\rho+\frac{\epsilon+p}{c^2} +\frac{b^2}{4\pi c^2}\right)u^\mu u^\nu \\
  %%%+\left(p+\frac{b^2}{8\pi}\right)g^{\mu\nu}-\frac{b^\mu b^\nu}{4\pi}
  %%% Heaviside-Lorentz units
  T^{\mu\nu}=\left(\rho+\frac{\epsilon+p+b^2}{c^2}\right)u^\mu u^\nu %\\
  +\left(p+\frac{b^2}{2}\right)g^{\mu\nu}-b^\mu b^\nu
\end{multline}
is the total energy-momentum tensor in Heaviside-Lorentz units; $\epsilon$, $\rho$, and $p$ are the fluid-frame internal-energy density, mass density, and gas pressure; $u^\mu$ is the fluid four-velocity; $b^\mu$ is the fluid-frame magnetic four-vector; $\tensor[^*]{F}{^{\mu \nu}}=b^{\mu}u^{\nu}-b^{\nu}u^{\mu}$ is the dual of the electromagnetic field tensor; and $g_{\mu\nu}$ is the four-dimensional metric tensor. The equation of state is that of a perfect gas with an adiabatic index of 4/3.

It is useful to introduce a foliation to divide spacetime into three-dimensional spatial hypersurfaces threaded by a universal time coordinate \citep{1982MNRAS.198..339T}. The spatial hypersurfaces are normal to a chosen field of observers $n_\mu$. We often represent the magnetic field using the three-dimensional vector measured by these fiducial observers, $B^i = - \tensor[^*]{F}{^{i \nu}} n_\nu$ \citep[e.g.][]{2004MNRAS.350..427K}. The three-dimensional metric tensor on the spatial hypersurfaces is indicated by $\gamma_{ij}$.

In order to follow the evolution of the accreting gas and separate it from the tenuous background magnetosphere we add a new quantity, $\mathcal{F}$, to the ideal GRMHD system; this auxiliary field is evolved as a passive scalar,
\begin{equation}
    \nabla_\mu (\mathcal{F} \rho u^\mu) = 0.
    \label{eq:passivescalar}
\end{equation}

\subsection{Coordinates}

The simulations are performed in spherical coordinates $(r,\theta,\phi)$ over the ranges $r = [\rNS, \rout]$, $\theta = [0,\pi]$, and $\phi = [0,2\pi]$. We set the radius of the neutron star to $\rNS = 4\,\rg$, where $\rg = G\MNS/c^2$ is the star's gravitational radius, and take the grid's outer radius to be $\rout = 3\times 10^3\,\rg$. %Unless explicitly indicated otherwise, we use the convention that $G = \MNS = 1$. %%% $G = \MNS = c = 1$. 

Near the star the cells' radial depth $\Delta r$ increases linearly with the radial coordinate  $r$, giving cells with approximately constant aspect ratio in the $r$--$\theta$ plane; beyond $r = 120\rgu$ their radial depth increases more rapidly. A two-dimensional coordinate mapping is applied in the $r$--$\theta$ plane, giving cells with roughly constant meridional extent $\Delta \theta$ near the star while focusing resolution toward the equator beyond $r \sim 20\,\rg$. The cells both near the polar axis and close to the inner boundary are mildly stretched away from the poles; this ``cylindrification'' increases the maximum stable time step. These coordinate transformations are described in \citet{2017MNRAS.467.3604R}.

\subsection{The star}
The star rotates with angular velocity $\Omegastar = 0.05\crg$, which places its light cylinder at $\RLC = c/\Omegastar = 20\,\rg$. We use the Kerr spacetime in the Boyer-Lindquist foliation, and set the Kerr spin parameter to $a = 1/3$, which is approximately appropriate for the star's rotation rate. The Keplerian angular velocity is $\Omega_{\rm K}(r) = c/(r\sqrt{r/\rg} + a\rg)$ for the prograde orbits of relevance here, placing the corotation radius, at which the stellar spin and Keplerian angular velocities are equal, at $\rco = \left(\rlc/\rg - a\right)^{2/3}\rgu \approx 7.29\rgu$.

The boundary conditions at $r = \rNS$ allow accreting material to fall through the stellar surface and, where accreting gas is not present, a magnetically dominated pulsar wind to be driven. An outflow condition is used at $r=\rout$. These boundary conditions are described fully in Appendix~\ref{sec:app_BCs}. Transmissive boundary conditions are applied along the coordinate axis, $\theta = 0, \pi$: a cell at $\phi = \phi_0$ reaches across the pole for its neighboring state to the cell at $\phi = \phi_0 + \pi$. Periodic boundary conditions are applied in the azimuthal $\phi$ direction.

\subsection{The torus}
The simulations are initialized with an equilibrium torus \citep{1976ApJ...207..962F,1985ApJ...288....1C} that acts as the gas reservoir for the accretion flow. The torus has its inner edge at $\rinner = 40\,\rg$ and pressure maximum at $\rmax = 60\,\rg$, at which point the mass density is $\rho_{\rm max}$. The torus gas rotates in the same direction as the star and its specific angular momentum, $l = - u_\phi/u_t$, scales with the von Zeipel parameter $\lambda = \sqrt{l/\Omega}$ as $l = c\, \lambda^{1/3}$, where $\Omega = u^\phi/u^t$ is the angular velocity. A magnetic field is added to the torus with vector potential $A_\phi \propto r^4 \rho^2$, forming one large loop of nested flux surfaces, and is normalized such that $\beta_{\rm max} \equiv 2 \, {\rm MAX}(p)/{\rm MAX}(b^2) = 100$, giving plasma $\beta \gtrsim 100$ everywhere. Random pressure perturbations are added at the $4 \times 10^{-2}$ relative level to encourage the development of turbulence.  

\subsection{The magnetosphere}
The initial stellar magnetic field is set using the azimuthal vector potential component for a potential dipole in the Schwarzschild metric \citep{1983ApJ...265.1036W},
\begin{equation}
    A_{\phi, \rm WS}(r,\theta) = \frac{3 \mu \sin^2\theta}{2\MNS} \left[ x^2 \ln(1-x^{-1}) + x + \frac12\right],
    \label{eq:WassShap}
\end{equation}
where $x \equiv r/2\MNS$. Here $\mu$ is the star's magnetic dipole moment, which is effectively in units\footnote{The factor of $4\pi$ comes from the denominator of Equation~(\ref{eq:WassShap}) in Heaviside-Lorentz units; i.e.\ the true Heaviside-Lorentz magnetic moment is larger than the value we quote by $4\pi$.} of $4\pi\rg^3 \sqrt{\rho_{\rm max}c^2}$. Henceforth the density and magnetic moment will be quoted without units, and are implicitly in units of $\rho_{\rm max}$ and $4\pi\rg^3 \sqrt{\rho_{\rm max}c^2}$ respectively. In this work the star's dipole moment vector and angular velocity vector are always either exactly parallel or antiparallel. We defer an investigation of the effects of obliquity between spin and magnetic axes to future work; see \citet{2012MNRAS.421...63R,2013MNRAS.430..699R,2021MNRAS.506..372R} for non-relativistic studies of accretion onto oblique rotators. The magnetization near the surface is generally $\gtrsim 10^4$ in the absence of accreting material.

We deform the stellar magnetic field to flow around the torus, so that no field lines couple the star and the torus in the initial conditions; see Appendix~{\ref{sec:deform}} for details. This is appropriate, since we place the torus beyond the star's light cylinder where accreting gas incoming from large radii should be initially unconnected to the star. There are three further practical benefits: (i) any coupling between the star and accretion flow that develops can then be attributed to simulated physical processes, rather than being ``baked in'' through the initial conditions; (ii) excluding the stellar field from the torus permits a cleaner investigation of the effect of changing the relative orientation of the stellar and torus magnetic fields, as otherwise the two components' reinforcement or cancellation leads to different magnetic field distributions throughout the initial torus; (iii) because we wrap around the torus those field lines which will be opened to infinity by the star's rotation at the beginning of the simulation, this step prevents the formation of a blanket of disconnected field lines trapped between the star and the accretion flow.

\subsection{Simulation parameters}
We investigated four values of the stellar magnetic moment, $\mu = \{5, 10, 20, 40\}$. For each, we performed two simulations in which the inner stellar dipole and torus field loop were either parallel or antiparallel where they met at the equator. Our principal set of simulations were performed at a resolution of $N_r \times N_\theta \times N_\phi = 384 \times 256 \times 128$ cells in each direction. In addition we ran two simulations at $\mu=10$, for both relative star--torus field orientations, at higher resolution in the toroidal direction, $N_\phi^{\rm high} = 256$. 
The eight standard-resolution simulations will be referred to with a shorthand specifying the magnetic moment and star--torus field orientation, for example 20--parallel or 5--antiparallel, while an additional postfix will indicate the two higher-resolution runs, as in 10--parallel--\nphihi. 
These ten primary simulations were evolved for 3.5--4$\times 10^4\rgc$, roughly 13 Keplerian orbits at the torus's pressure maximum or 300 stellar spin periods ($\Pstar = 2\pi \RLC/c \approx 126 \rgc$). Each combination of magnetic moment and relative field orientation was also studied at several lower grid resolutions, with broadly similar outcomes.

The numerical scheme employed second-order reconstruction, second-order predictor--corrector time advancement, Lax-Friedrichs fluxes, and a monotonized-central limiter with a steepener of 1.95.

\section{Results}
\label{sec:results}

\subsection{General behavior}
\label{sec:general}

We first focus on a single simulation, 10--antiparallel--\nphihi. From $t = 0$ the non-rotating stellar magnetosphere relaxes toward a nearby steady state. Stellar rotation is introduced\footnote{The spacetime's Kerr spin parameter is held constant for the entirety of the simulation.} at $t=500\rgc$, increasing linearly to its final steady angular velocity over $\Delta t = 40\rgc$. Rotation of the stellar surface launches Alfv\'{e}n waves into the magnetosphere and quickly produces a configuration very similar to that of an isolated pulsar: around the equator there is an approximately dipolar closed magnetosphere extending to the light cylinder, while field lines emerging from nearer the star's poles are open to infinity, flowing around the torus. We use a subscript ``0'' to label quantities referencing this pre-accretion state: the total open magnetic flux is $\ffo$, which extracts stellar rotational energy at rate $\Lo$, applying a torque $\No$ to the star\footnote{Because the torus is initially well beyond the light cylinder the values of open flux, extracted power, and torque are very similar to what is found when the star is truly isolated, with no torus. We will use $\ffo$, $\Lo$, and $\No$ to refer to both scenarios.}; we adopt the convention that $\No > 0$ implies spin-up of the star.

Meanwhile, magnetic stresses begin to disrupt the torus. Shear along field lines causes angular momentum transport and eventually the onset of the magnetorotational instability \citep[MRI;][]{1998RvMP...70....1B,2003ApJ...592.1060D}. The torus material begins to move inward at $t \sim 3,\!000\rgc$ and enters the star's light cylinder at $t \sim 5,\!000\rgc$. The accretion flow quickly reaches the inner magnetosphere, where its inward movement is halted at the magnetospheric radius, $\rmag \sim 6.5\,\rg$, by the closed zone's magnetic pressure. As we are evolving a total-energy conservation equation, dissipation from shocks and MRI-generated turbulence causes the accretion flow to puff up into a thick disk. 

Of the stellar magnetic flux that initially closed between $\rmag$ and $\RLC$, some enters the disk and some opens up, adding to the original rotationally opened flux. The star's electromagnetic ``pulsar wind'', strengthened by the newly opened flux, is collimated somewhat by the thick accretion flow, forming what is in effect a Poynting-flux-dominated relativistic jet along the star's spin axis. 

\begin{figure*}
    \centering
\includegraphics[width=\textwidth]{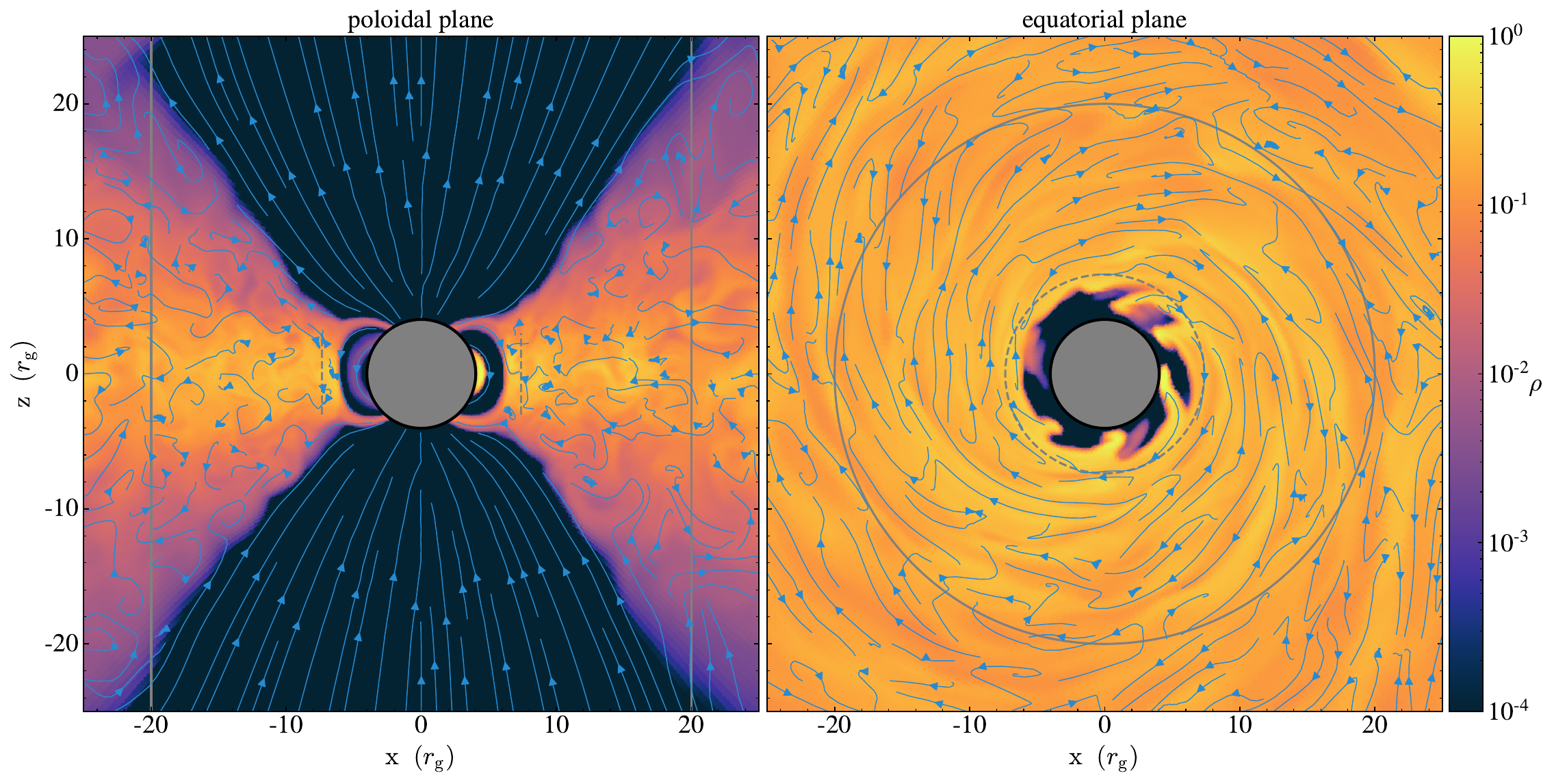}
    \caption{Poloidal ($x$--$z$) and equatorial ($x$--$y$) slices of the high-resolution $\mu=10$ simulation, 10-antiparallel-\nphihi, at $t = 11,560\,\rg/c$. The color represents the mass density $\rho$, while blue line segments indicate the direction of the in-plane component of the magnetic field. The star's light cylinder $\RLC$ and corotation radius $\rco$ are marked by solid and dashed gray lines respectively. Several interchange streams can be seen spiraling inward from the magnetospheric boundary at $\rmag \sim 6.5\rgu$.}
    \label{fig:mu10_pol_eq}
\end{figure*}

\begin{figure}
    \centering
\includegraphics[width=\columnwidth]{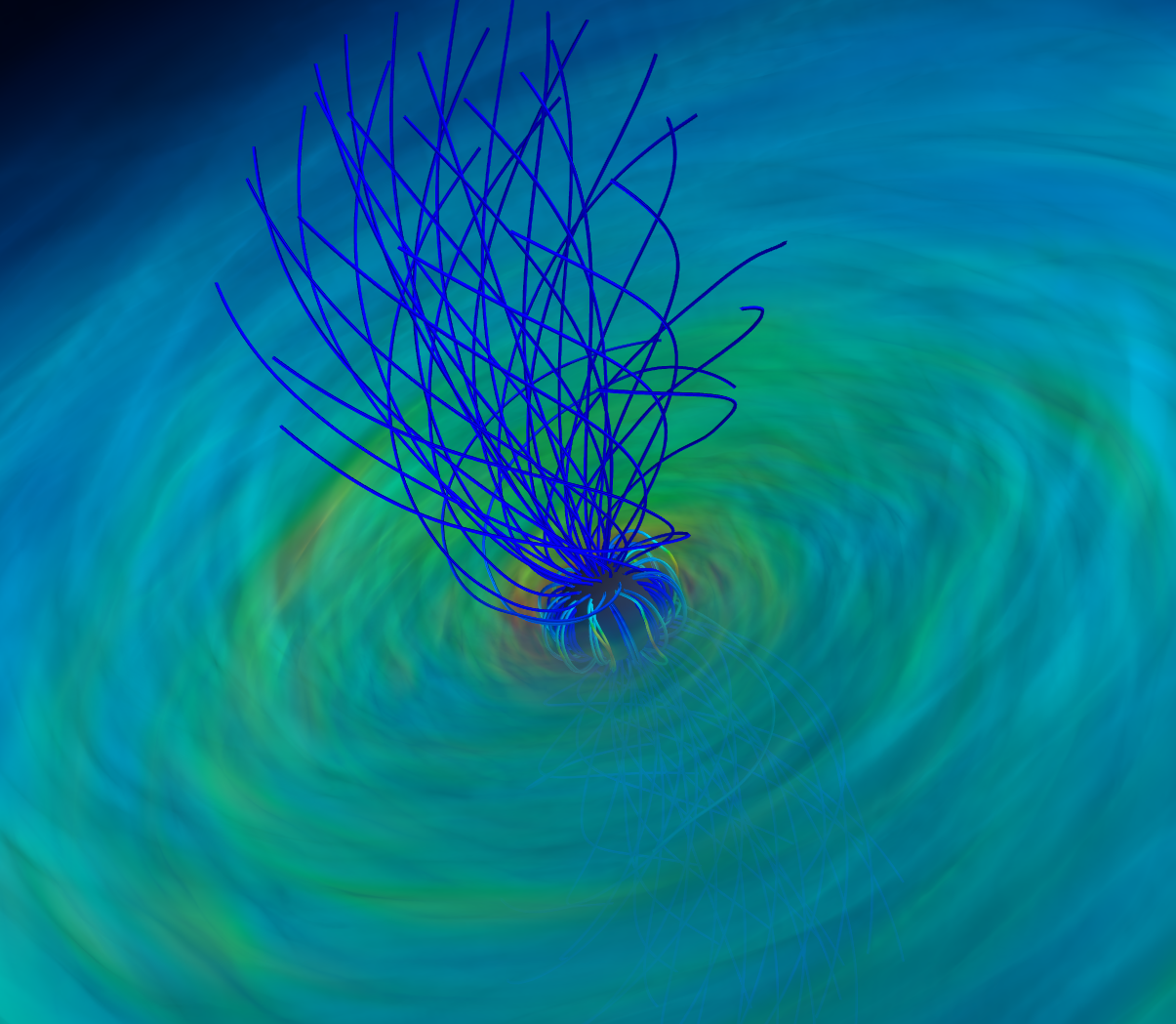}
    \caption{Volume rendering of the matter density field in 10--antiparallel--\nphihi with 3D stellar magnetic field lines colored by the local density, increasing in both cases from dark blue to red. The closed field lines are deformed and displaced by the impinging interchange streams, while the open field lines are swept back into a smooth large-scale helix.}
    \label{fig:volume3D}
\end{figure}

Figure~\ref{fig:mu10_pol_eq} shows illustrative poloidal- and equatorial-plane slices of this simulation. There is a sharp distinction between the matter-dominated accretion flow and the magnetically dominated (i.e.\ very low density) inner magnetosphere and jet funnel. In the turbulent accretion flow the magnetic field is disordered, with longer structures in the azimuthal direction due to the stretching effect of nearly Keplerian rotation, while in the jet region the magnetic field is twisted into a helix (see Figure~\ref{fig:volume3D}), appearing nearly radial in poloidal cross-section.

In this simulation the magnetospheric boundary lies inside the corotation point ($\rmag < \rco$) and disk material does not experience a centrifugal barrier to accreting onto the star.
At the magnetospheric radius, accreting material is directed by the star's magnetic field into thin accretion columns, through which the gas reaches the stellar surface and passes smoothly across the simulation's inner boundary. Here, because the star's spin and magnetic axes are aligned, the accretion columns form quasi-axisymmetric curtains; for general rotators with non-zero spin--magnetic obliquity two distinct streams would form, one for each pole. 

The magnetospheric boundary is unstable to non-axisymmetric modes of the interchange (magnetic Rayleigh-Taylor) instability, in which thin streams of the accreting material push aside the star's dipole-like closed field lines and move inward \citep{1976ApJ...207..914A,1980ApJ...235.1016A,1992ApJ...386...83K,1995MNRAS.275.1223S}. This behavior has been observed in non-relativistic simulations \citep{2008MNRAS.386..673K,2016MNRAS.459.2354B,2022ApJ...941...73T,2023arXiv230915318Z}. Several interchange streams are visible in Figure~\ref{fig:mu10_pol_eq}; in this simulation we see up to seven or eight streams at a time. Some of them reach the stellar surface, leading to matter accretion near the equator. This interchange-mode accretion occurs concurrently with accretion through the columns, which remain quasi-steady.

\begin{figure}
    \centering
    \includegraphics[width=\columnwidth]{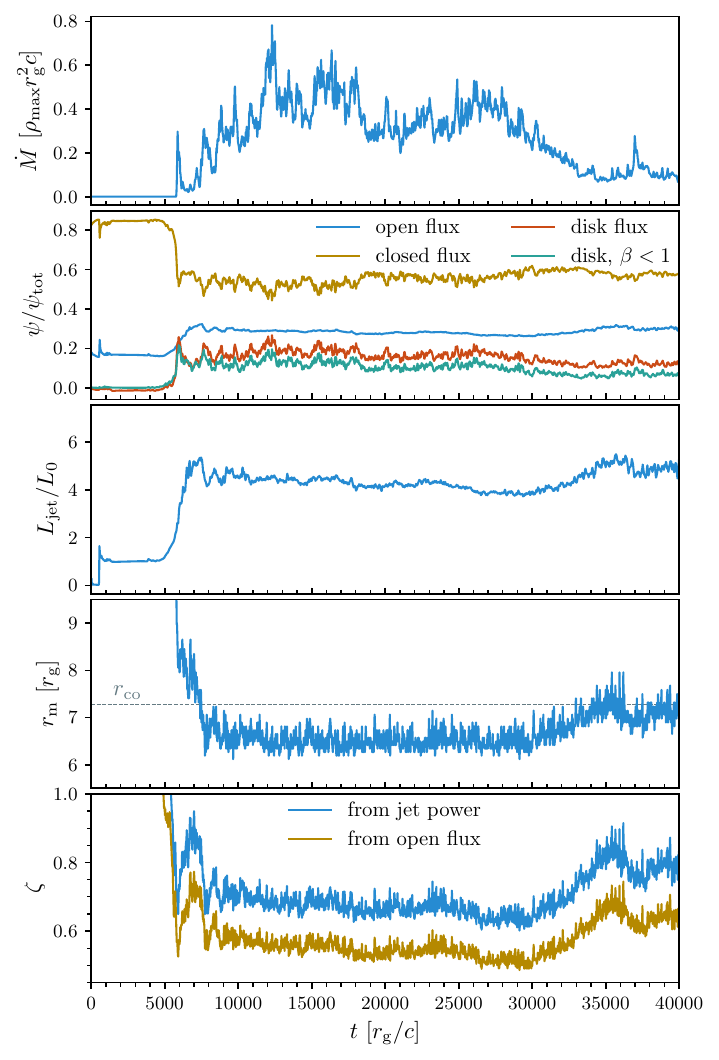}
    \caption{Derived quantities versus time for the 10--antiparallel--\nphihi simulation. From top: mass accretion rate onto the star; open, closed, and disk-connected magnetic flux in units of the total stellar flux $\fftot$; electromagnetic jet power $\Lj$ in units of the isolated pulsar wind power $\Lo$; magnetospheric radius $\rmag$, with a horizontal line for the corotation radius $\rco$; flux-opening efficiency factor $\zeta$, derived both from the jet power and the open flux.}
    \label{fig:10ar4_multivar_time}
\end{figure}

Figure~\ref{fig:10ar4_multivar_time} shows several of this simulation's global quantities as a function of time. The mass accretion rate $\mdot$ is measured over the entire stellar surface. Once some disk material has passed through the corotation radius there is a gradual build up in accretion rate over $\Delta t \sim 7,\!000\rgc$, after which it remains approximately steady until $t \sim 30,\!000\rgc$. From that point on the accretion rate declines, as the torus material has become depleted through accretion, winds, and outflows.  

The amount of open and closed magnetic flux is shown in the second panel, normalized by the total absolute flux through the stellar surface, 
\begin{equation}
\fftot = \int_0^{2\pi}\int_0^\pi |B^r(\rNS,\theta,\phi)| \sqrt{\gamma}\, {\rm d}\theta\, {\rm d}\phi.
\end{equation}
This integral counts each field line twice. The open flux is defined as the magnetic flux through a sphere having the light-cylinder radius, $r=\RLC$, in the $\sigma > 1$ regions around the poles; the magnetization 
\begin{equation}
 \sigma = \frac{b^2}{\rho c^2 + p + \epsilon}
\end{equation}
is the ratio of magnetic and hydrodynamic enthalpy densities. We use $\sigma$ to distinguish magnetically from materially dominated regions, and generally we define the disk as having $\sigma < 1$ and the magnetosphere or jet as having $\sigma > 1$. The closed flux is twice the magnetic flux through the equatorial plane where $\sigma > 1$ inside the light cylinder. These definitions of open and closed flux appear to be reliable proxies for field lines of the desired connectivity. The disk-connected flux is the remainder: $\ff_{\rm disk} = \fftot - \ffopen - \ff_{\rm closed}$; this flux enters the accretion flow inside the light cylinder, and can remain connected to the disk for short or long periods. Finally, we show a curve for the flux connected to the disk in $\beta < 1$ regions, in other words flux passing through the equator where $\sigma < 1$ but also $\beta < 1$. These field lines constitute an intermediate region where both magnetic and hydrodynamic forces are important. 

The closed flux rapidly drops once the accretion flow enters the light cylinder, with the open and disk-connected fluxes rising equally quickly. The opening process can proceed easily in this case, as the arriving stellar field lines are antiparallel to the star's closed magnetosphere, allowing immediate magnetic reconnection. The open flux stays fairly stable for the remainder of the simulation, with more transference occurring between the closed and disk-connected zones. Of the flux that enters the accretion flow, most of it remains in regions where magnetic forces are strong ($\beta<1$), despite these strong-magnetic-pressure regions making up only roughly 17\% of the disk volume inside the light cylinder (see Section~\ref{sec:afprops}).

The relativistic jet power is associated with the integrated flux of electromagnetic energy at infinity, through a sphere at $r=\RLC$ in the polar $\sigma > 1$ ``jet funnel'' regions,
\begin{equation}
    \Lj = -\int \int \left(b^2 u^r u_t - b^r b_t\right) \sqrt{-g}\, {\rm d}\theta\, {\rm d}\phi.
\end{equation}
The jet power rises with the increase in open magnetic flux, as generally $\Lj \propto \ffopen^2$. The collimating effect of the disk, which pushes the open field lines away from the equator, also increases the energy (and angular momentum) extracted per unit open magnetic flux. The jet power is stable for most of the steady-state part of the simulation. 

The disk--magnetosphere boundary is generally highly non-axisymmetric. We define an effective magnetospheric radius $\rmag$ as the point at which the magnetization at the equator, averaged azimuthally and between $\theta = \pi/2 \pm \pi/24$, has decreased to $\sigma =1$. This is biased toward giving an underestimate of $\rmag$, since it includes incoming magnetic Rayleigh-Taylor streams that are properly lying inside the magnetospheric radius. In this simulation, $\rmag$ moves inward at roughly the same rate as $\ffopen$ increases, eventually stabilizing at $\rmag \sim 6.5\rgu$ with a standard deviation of $0.2\rgu$. 

The final panel of Figure~\ref{fig:10ar4_multivar_time} shows the flux-opening efficiency parameter $\zeta$. One can construct a simple model for the open flux of an accreting pulsar in terms of that of the equivalent isolated pulsar $\ffo$,
\begin{equation}
    \ffopen = \zeta \frac{\RLC}{\rmag}\, \ffo
    \label{eq:ffmodel}
\end{equation}
when $\rmag < \RLC$ \citep{2016ApJ...822...33P}. Here $\zeta = 1$ corresponds to perfect opening of all of the previously closed flux between $\rmag$ and $\RLC$, while $\zeta < 1$ quantifies how much of this disk-interacting flux remains outside the open-flux region. We can use this model for $\ffopen$ to estimate the star's jet power,
\begin{equation}
    \Lj = \zeta^2 \left( \frac{\RLC}{\rmag}\right)^2 \Lo, 
    \label{eq:Ljmodel}
\end{equation}
or equivalently its spin-down torque; the spin-down power for an isolated aligned rotator in flat spacetime is given by $\Lo = \mu^2 \Omegastar^4/c^3$ \citep{2005PhRvL..94b1101G,2006ApJ...648L..51S}. We can use Equations~(\ref{eq:ffmodel}) and (\ref{eq:Ljmodel}) to infer effective $\zeta$ values, $\zeta_{\ff} = (\ffopen/\ffo)\, (\rmag/\RLC)$ and $\zeta_{\rm jet} = \sqrt{\Lj/\Lo}\, (\rmag/\RLC)$ respectively, as we can measure some quantities ($\ffopen$, $\Lj$, $\rmag$) and the rest are known input parameters. 

In this simulation the steady-state efficiencies are $\zeta_{\rm jet} \sim 0.69$ and $\zeta_\ff \sim 0.57$. There is a consistent multiplicative factor of $\sim 1.22$ between the two values, which may be due to the disk's collimation causing the energy extracted on open field lines to increase faster than $1/\rmag^2$ with decreasing $\rmag$.

Near the end of the simulation the accretion flow has lost much of its mass to accretion and outflows; the effective accretion rate supplied to the magnetospheric boundary decreases, resulting in a decrease in $\mdot$ onto the star and the shifting outward of the magnetospheric radius. Interestingly, the opening efficiency $\zeta$ increases, causing the amount of open flux, and consequently the jet power, to also increase, as can be seen in Figure~\ref{fig:10ar4_multivar_time}. Over this period there is therefore an anti-correlation between $\mdot$ and $\Lj$ which runs counter to the basic idea that a deeper-penetrating disk opens more flux and induces a more powerful jet. We will return to this behavior when we discuss flux opening in Section~\ref{sec:interchange}.

\subsection{Effect of the star--disk relative field orientation}

The magnetic field advected inward by the accretion flow can make any angle with the star's dipole field, and in general this angle will depend on both $\phi$ and $t$. In a simulation the average angle will retain a memory of the initial conditions. We investigate the two extreme possibilities: the initial field in the torus is either parallel or antiparallel to the star's closed field lines. In axisymmetric simulations the two cases produce very different behavior, because significant reconnection between stellar and disk fields only occurs with the antiparallel choice, and the flux surfaces are unable to change their overall orientation. Stellar field lines open easily via reconnection in the antiparallel scenario, giving a strong relativistic jet, whereas when the two fields are parallel the field lines close up and the jet is suppressed (see \hyperlink{PT17target}{PT17}).  

One may expect that this dichotomy would be much less pronounced in 3D, because (a) individual disk field lines can change their orientation by twisting around in the azimuthal direction, and (b) the stellar field can enter the disk through non-axisymmetric fingers and then be opened by angular velocity shear along the field lines  (see Section~\ref{sec:interchange}).

\begin{figure}[tb]
    \centering
    \includegraphics[width=\columnwidth]{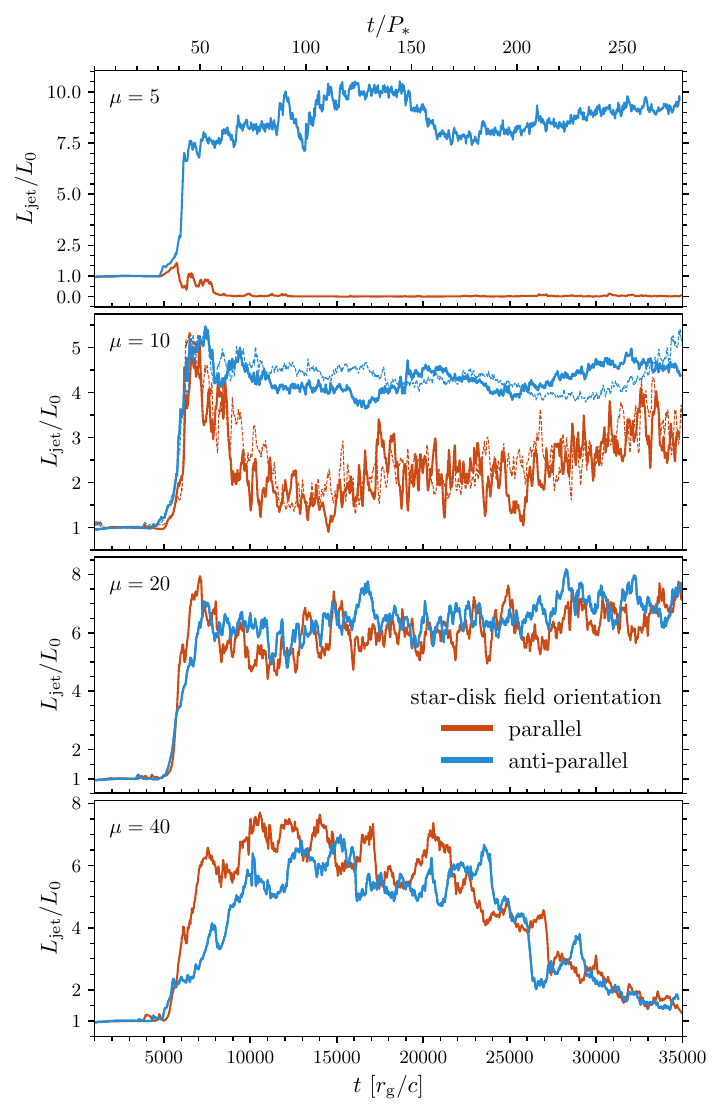}
    \caption{Electromagnetic jet power as a function of time for the four magnetic moment values.  Color indicates the relative orientation of the stellar and torus magnetic field in the initial conditions: red for parallel, blue for antiparallel. Thin dotted lines indicate the two $\mu=10$ high-$\phi$-resolution simulations.}
    \label{fig:Ljet_time}
\end{figure}

Figure~\ref{fig:Ljet_time} shows the relativistic jet power for both orientation choices, for each of our magnetic moment values. At low magnetic moment (equivalently, high accretion rate), $\mu = 5$, the strong distinction between parallel and antiparallel initial orientations is retained --- the antiparallel simulation shows a strong jet that is significantly more powerful than the original pulsar wind, while the jet power in the parallel run is suppressed well below $\Lo$ and goes almost to zero. On the other hand, for higher stellar field strengths, $\mu = 20$ or 40, the effect of the orientation on the jet power disappears nearly completely, with both choices giving powerful jets. 
 
The intermediate magnetic moment, $\mu = 10$, interpolates between the two scenarios. When the disk first enters the light cylinder, both orientations see the jet power rising rapidly to the same peak, $\Lj \sim 5.5\, \Lo$. However, the power remains approximately steady at that level in the antiparallel simulation, while in the parallel one it drops nearly equally rapidly, falling to $\sim 2\, \Lo$ by $t = 10,000 \rgc$. From that point it slowly recovers, approaching the antiparallel-run jet power by the end of the simulation. The jet power is much more variable in the parallel simulation, with a standard deviation of $0.35\,\Lo$ between $t=10,000$ and $15,000\rgc$, as compared to $0.12\,\Lo$ for the antiparallel case. The magnetospheric radius is also more variable in the parallel simulation, with a standard deviation of $0.38\rgu$ as opposed to $0.22\rgu$ over the same period. 

The jet power for the higher-$\phi$-resolution $\mu = 10$ simulations is also shown in Figure~\ref{fig:Ljet_time}, with thin dotted lines. The overall behavior is very similar, as are the various measurements for the variability in $\Lj$ and $\rmag$ quoted above. Our results appear to be insensitive to azimuthal resolution.

{\color{red} }

\begin{figure*}
    \centering
    \includegraphics[width=\textwidth]{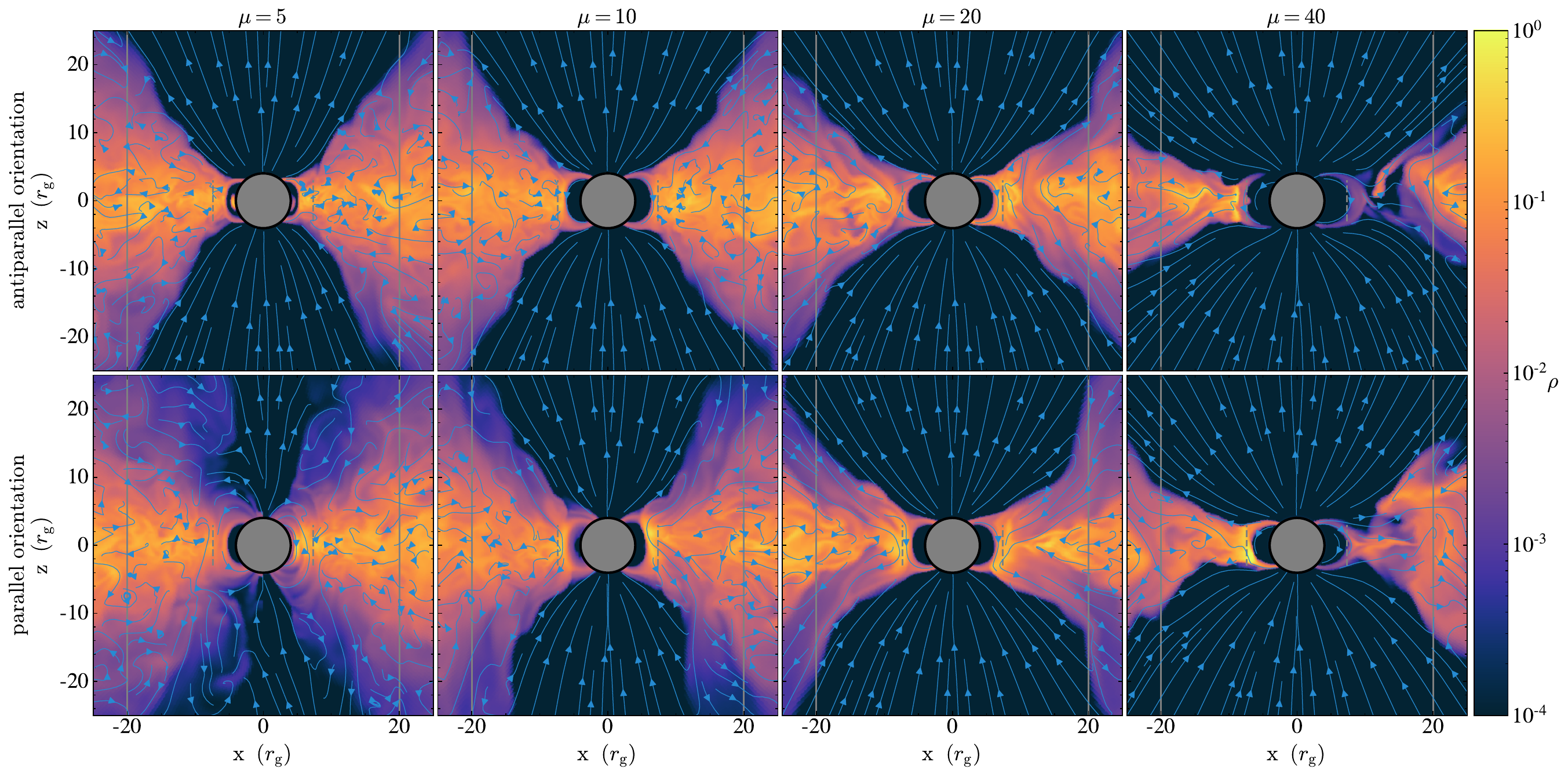}
    \caption{Poloidal slices of the mass density and in-plane magnetic field at $t \sim 10,500 \rgc$, for each combination of the stellar magnetic moment and star--disk field orientation. The initial field orientation becomes progressively less important as $\mu$ increases (or, equivalently, as the accretion rate decreases).}
    \label{fig:mu5-40_poloidal}
\end{figure*}

Figure~\ref{fig:mu5-40_poloidal} shows representative snapshots of each simulation's poloidal-plane structure. There is a clear difference between the orientations at $\mu = 5$: the antiparallel simulation has thin accretion columns and a large, clean jet funnel, while in the parallel run the disk material nearly envelopes the star, with accretion proceeding via thick columns and disk material nearly reaching the poles, choking off the relativistic jet. The two cases look much more similar at $\mu = 10$ and 20, but in the parallel runs one can see field lines connecting the star to the disk's outer layers and inflating outward in a banana shape, whereas there is very little star--disk coupling in the antiparallel simulations.

The $\mu = 40$ simulations are in a different regime. The accretion flow is mostly kept at or beyond the corotation radius, and so the centrifugal barrier prevents gas from accreting onto the star. The stellar electromagnetic wind or jet pushes outward against the disk, and any disk material that becomes connected to closed or open stellar field lines is accelerated in the azimuthal direction, increasing its angular momentum. These effects combine to push the accretion flow outward in what is termed the propeller state. This results in a disrupted, highly dynamic, low-density disk. When the accretion flow remains near the corotation radius, as in these simulations, accretion can still occur intermittently and some material can persist in the inner magnetosphere even when most of the disk has been pushed away (see for example the 40--parallel panel in Figure~\ref{fig:mu5-40_poloidal}).

\begin{figure*}
    \centering
    \includegraphics[width=\textwidth]{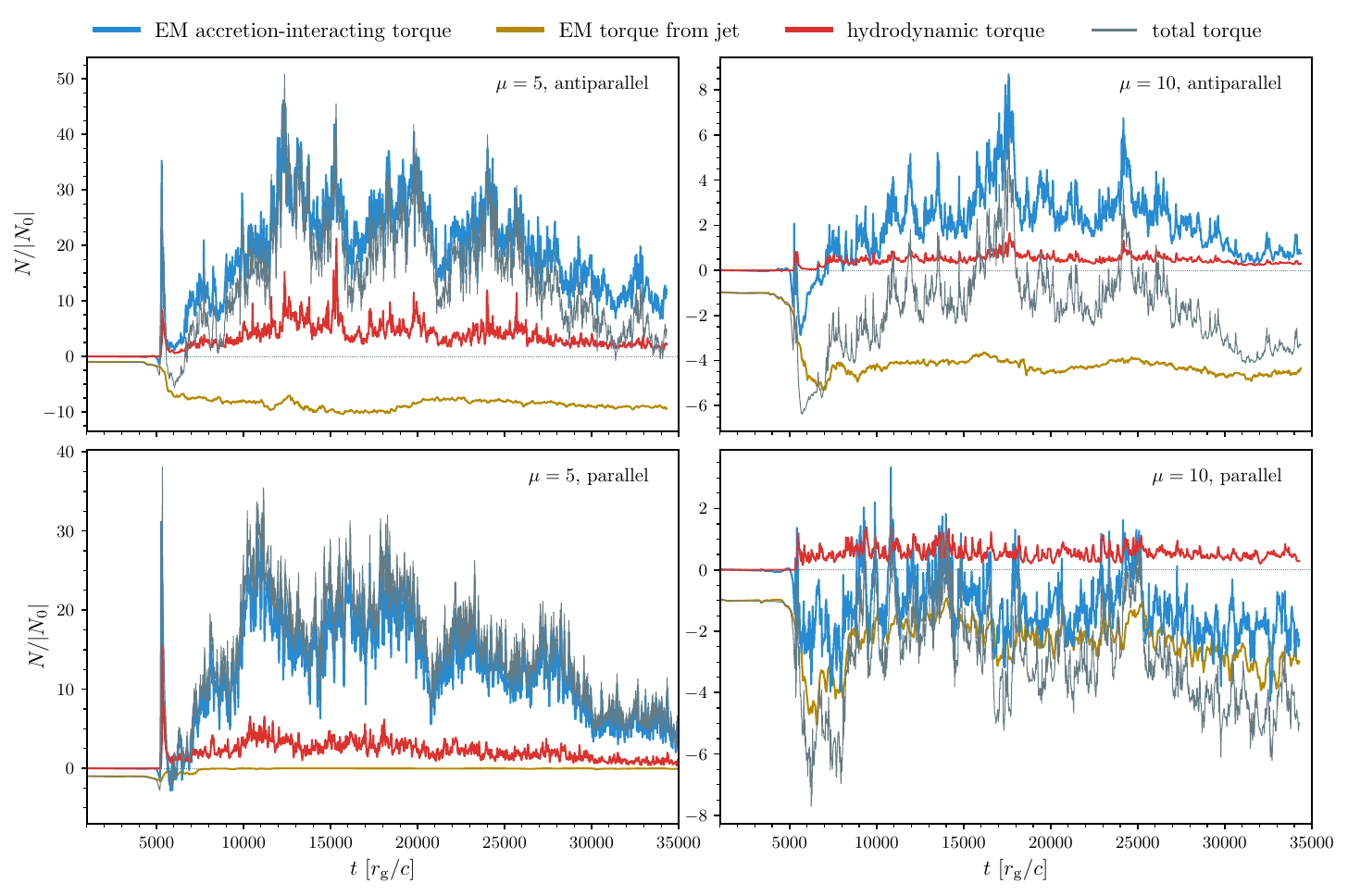}
    \caption{Electromagnetic and hydrodynamic contributions to the torque on the star in the $\mu=5$ and 10 simulations, in units of the isolated-pulsar torque $\No$. The ``accretion-interacting'' torque is applied by all field lines that couple to the accretion flow, both inside and outside the columns. The total torque is the sum of the three components. Positive (negative) values indicate spin-up (spin-down) of the star. The $\mu=5$ simulations spin-up consistently, while the $\mu = 10$ runs display brief, intermittent torque reversals.}
    \label{fig:torques}
\end{figure*}

The torque $N$ on the star is the negative of the outward flux of angular momentum at infinity; its electromagnetic (EM) and hydrodynamic contributions are given by
\begin{subequations}
\begin{align}
\Nem    &= - \iint \left( b^2 u^r u_\phi - b^r  b_\phi \right) \sqrt{-g}\, {\rm d}\theta\, {\rm d}\phi, \label{eq:Nem}\\
\Nhydro &= - \iint \left( \rho + \frac{p + \epsilon}{c^2}\right) u^r u_\phi \sqrt{-g}\, {\rm d}\theta\, {\rm d}\phi. \label{eq:Nhydro}
\end{align}
\end{subequations}
Figure~\ref{fig:torques} shows the torque contributions for the $\mu = 5$ and 10 simulations. For the total electromagnetic and hydrodynamic torques the integral is taken over the entire stellar surface. 
The EM torque from the jet is found by integrating at $r = \RLC$ over the $\sigma > 1$ (i.e.\ open flux) region; this closely approximates the jet torque applied to the star since nearly all of the field lines in this region connect to the star and angular momentum is conserved by the numerical scheme. 

The difference between the total and jet EM torques is contributed by those stellar field lines that interact with the accretion flow, including both those field lines in the accretion column and those that are in the force-free region at the star but couple to the disk outside $\rmag$. The column always spins the star up (positive torque) but the disk-entering field lines can make a contribution of either sign, and so the combined ``accretion-interacting'' EM torque shown in Figure~\ref{fig:torques} can be positive or negative. The hydrodynamic torque comes exclusively from material falling through the inner boundary in the accretion column and is always positive or zero.

At $\mu=5$ there is comparatively little difference between the parallel and antiparallel field orientations, despite these scenarios having radically different jet powers (Figure~\ref{fig:Ljet_time}) and overall field and matter distributions (Figure~\ref{fig:mu5-40_poloidal}). While the antiparallel run has a jet torque that is not present in the parallel case, the total torque is dominated by the large EM spin-up torque exerted by the columns and field lines interacting with the accretion flow inside $\rco$, where the disk rotates faster than the star. The hydrodynamic torque is subdominant but not insignificant.

At higher stellar field strength, $\mu=10$, the total EM torque is predominantly negative, causing the star to spin-down for most of the simulation. In the antiparallel run the EM accretion-interacting contribution is usually positive, indicating the dominance of torques from the column and field lines entering the inner disk, while in the parallel simulation it is generally negative due to more stellar field lines interacting with the disk outside the corotation radius. (We will return to this difference when we discuss flux opening in Section~\ref{sec:interchange}.) Both $\mu=10$ runs have brief, occasional periods in which the sign of the torque reverses and the star spins up. This behavior was also observed in the higher--$\phi$--resolution simulations, where the increased spin-up torque coincided with the moderate peaks in accretion rate seen in Figure~\ref{fig:10ar4_multivar_time}. 

For our highest magnetic moments, $\mu=20$ and 40, the hydrodynamic torque is very small, the jet torque is large, and the accretion-interacting EM torque is consistently negative for both orientations, resulting in strong spin-down (not shown in the figure).

\begin{figure}[tbh!]
    \centering
    \includegraphics[width=\columnwidth]{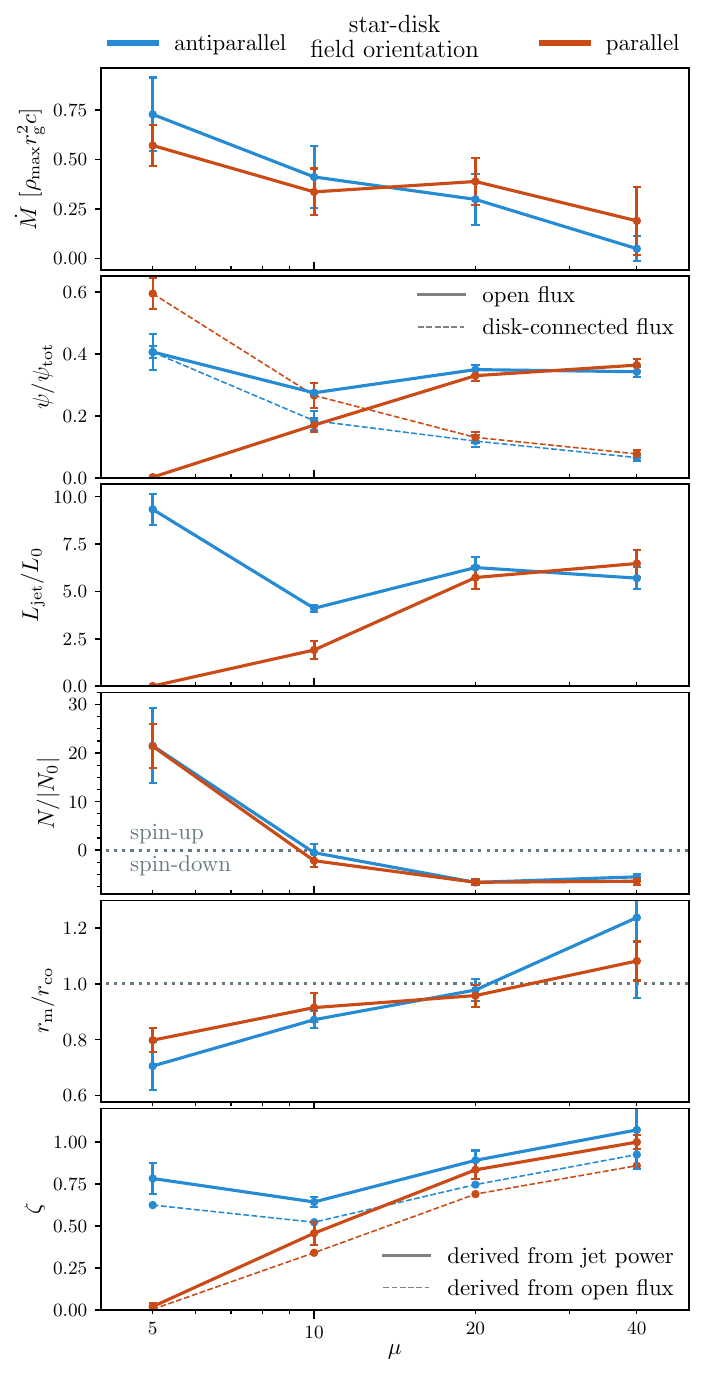}
    \caption{Quantities averaged over the steady-state period $\dtsteady$ for the eight combinations of relative field orientation and stellar magnetic moment. From top: stellar accretion rate; open and disk-connected stellar magnetic flux; electromagnetic jet power; stellar torque; magnetospheric radius in units of the corotation radius; flux-opening efficiency parameter ($\zeta_{\rm jet}$ and $\zeta_{\psi}$). Vertical bars represent the standard deviation over the averaging period. 
    }
    \label{fig:multivar_mu}
\end{figure}

Figure~\ref{fig:multivar_mu} presents global derived quantities averaged between $t = 10^4\rgc$ and $2\times 10^4\rgc$. This window of roughly 80 stellar spin periods, which we will refer to as $\dtsteady$, is the longest stretch over which all eight of our primary simulations are in an approximate steady state (see Figure~\ref{fig:Ljet_time}). Several of these quantities are insensitive to the initial star--torus relative field orientation over the full range of magnetic moments: the accretion rate onto the star, the torque applied to the star, and the location of the magnetospheric radius. The amount of magnetic flux coupling the star to the disk is weakly dependent on the orientation. On the other hand the amount of open magnetic flux, and hence the relativistic jet power and the flux-opening efficiency parameter, show a pronounced distinction between insensitivity to orientation at large $\mu$ and a very strong dependence at small $\mu$.

The star is closest to spin equilibrium (average $N\sim0$) at $\mu = 10$, as already suggested in Figure~\ref{fig:torques}. In these runs the magnetospheric boundary is well inside corotation, and the system is in the accreting regime. When $\rmag$ is closer to (though on average just inside) $\rco$, at $\mu = 20$, the star consistently spins down strongly, $N \approx -6.6 |\No|$. Due to the spin-down torques from the jet and field lines interacting with the outer accretion flow, spin equilibrium occurs when $\rmag < \rco$ rather than when these radii coincide.

The opening of stellar magnetic flux through interaction with the disk is usually efficient, particularly at larger $\rmag$ where nearly all field lines that touch the disk are opened ($\zeta \sim 1$). The antiparallel orientation consistently provides higher efficiency ($\zeta \gtrsim 0.75$), which may be expected given the ease of opening field lines through magnetic reconnection in this case. Only in the 5--parallel simulation, where the accretion flow reaches well past corotation and approaches  close to the surface, does the stellar field remain closed. 

The system is in the accreting state at $\mu = 10$ and the propeller regime at $\mu=40$, with an intermediate $\rmag \sim \rco$ state at $\mu = 20$. None of the system's averaged quantities change sharply as one passes between states. Accretion onto the star still occurs in these weak propeller simulations ($\rmag \gtrsim \rco$) because disk material occasionally enters through corotation. The flux-opening model for the relativistic jet, and the associated electromagnetic spin-down torque, remains valid in both accreting and propeller states.

\subsection{The interchange slingshot}
\label{sec:interchange}

\begin{figure*}
    \centering
    \includegraphics[width=\textwidth]{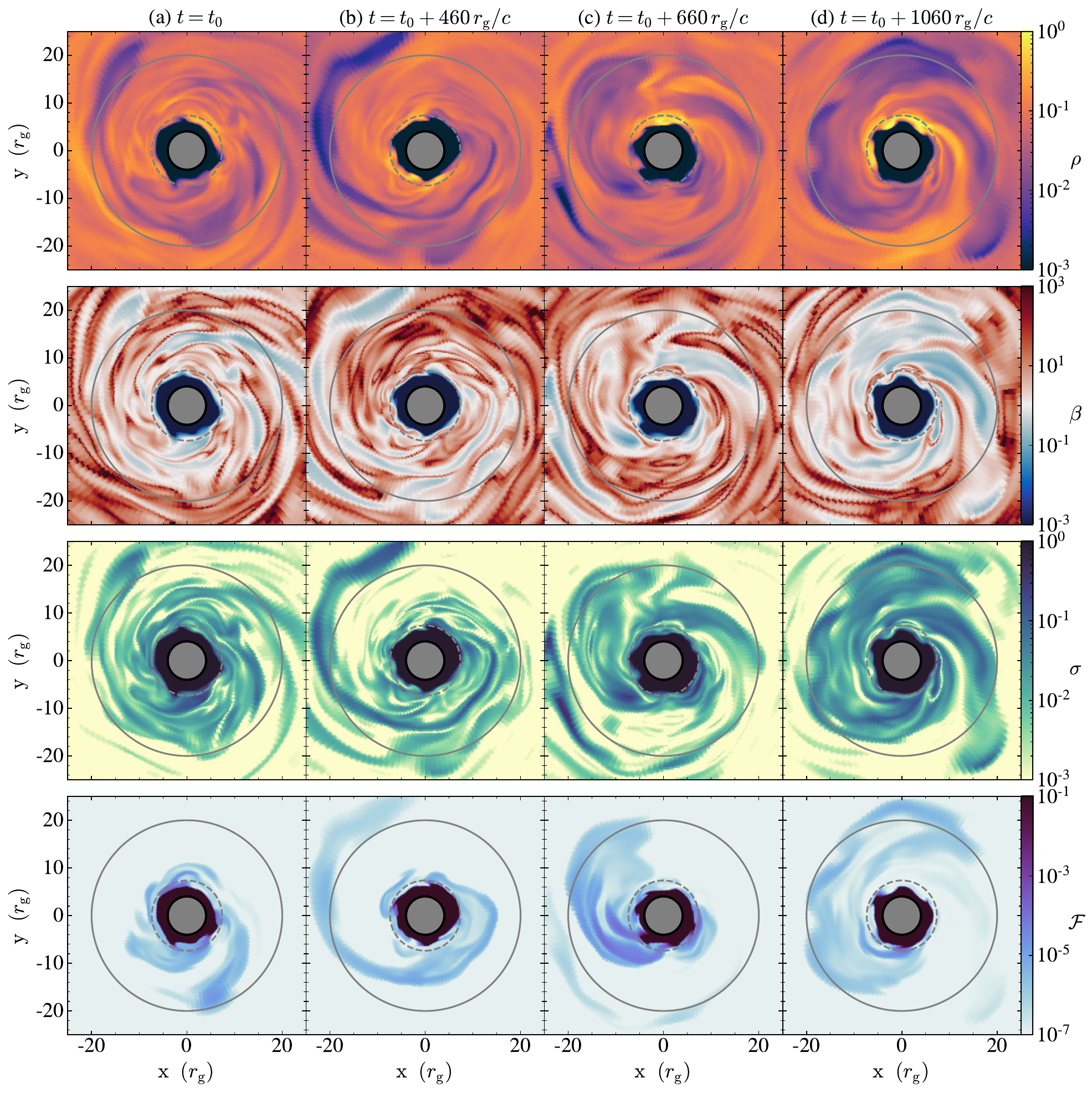}
    \caption{The interchange slingshot, from the 20--parallel simulation. Starting at $t_0 = 23,\!860\rgc$, several outward spirals of magnetically dominated material from the star's closed magnetosphere are produced over the next $\sim 8$ spin periods. The spirals are visible in density $\rho$, plasma $\beta$, and magnetization $\sigma$, but are clearest in the passive scalar $\mathcal{F}$ that directly traces material from the force-free magnetosphere. Equatorial slices are shown, with the light cylinder and corotation radius marked in gray.}
    \label{fig:centrifuge}
\end{figure*}

Why do the two relative field orientations produce radically different solutions at small magnetic moment (equivalently, high accretion rate) yet make almost no difference to most of the important large-scale properties when the star's magnetic field is stronger? These trends appear to be determined by whether the star's magnetic field is opened by the accretion flow or remains closed, with more flux coupling star and disk. 

It is straightforward to open the stellar field through magnetic reconnection when the inwardly advected disk field and the closed stellar field lines are roughly antiparallel where they meet: the disk's outermost field lines are now connected to the star, and are whipped into a rotating helix by the star's spin. This slings the disk material on these field lines outward, and they quickly join the existing bundle of magnetically dominated open field lines. A build up of magnetic pressure, rather than reconnection, occurs when the two flux systems are parallel at the interface. In axisymmetry this leads to flux \emph{closing} rather than opening in all cases (\hyperlink{PT17target}{PT17}); it is this relationship that is selectively broken in 3D.  

We propose the following mechanism, which we term the \emph{interchange slingshot}, that opens up closed stellar field lines when $\rmag \sim \rco$ or the system is in the propeller regime. A non-axisymmetric interchange perturbation develops at the magnetospheric boundary, with a low-density, magnetically dominated finger of the star's closed magnetosphere intruding into the disk. The finger, rotating at the stellar angular velocity $\Omegastar$, begins to interact with the disk gas orbiting at $\Omega_{\rm disk} \sim \Omega_{\rm K}(\rmag)$ at the same radius. 

\begin{description}[font=\normalfont]
\item[(a) $\rmag \ll \rco$] In this case $\Omegastar \ll \Omega_{\rm disk}$, and the slower finger pushes back on the disk gas behind it in rotational phase, reducing its angular momentum and causing it to spiral inward. As the disk material moves inward its angular velocity increases, and it catches and erases the magnetospheric finger.

\item[(b) $\rmag > \rco$] When $\Omegastar > \Omega_{\rm disk}$ the intruding finger pushes the disk gas ahead of it forward, increasing its angular momentum and leading it to spiral outward. The low-density finger can rapidly expand into the vacated space. A large plume of stellar magnetic field expands into the disk, forming a spiral as its angular velocity is progressively retarded by the ever slower nearby disk material. 

This results in a large velocity shear between the field lines' footpoints on the star and their locations in the plume, twisting the field lines in the azimuthal direction and building up a toroidal magnetic field component. The magnetic pressure of this toroidal field inflates the field lines in the vertical direction and eventually opens them entirely. The closed stellar field line splits into two parts: an open stellar field line and a disk field line (not connected to the star). 

\item[(c) $\rmag \lesssim \rco$] When the magnetospheric boundary is inside, but close to, the corotation point, the rate at which the finger is erased, as in (a), may be slower than the ordinary growth of the interchange instability. The perturbation grows until it reaches a point where $\Omega_{\rm disk} < \Omegastar$, and the accelerated expansion outlined above in (b) takes over.
\end{description}

In Figure~\ref{fig:centrifuge} one can see several spirals of magnetic field from the star's closed magnetosphere. They emanate from inroads that the inner magnetosphere makes into the accretion flow, extend beyond the corotation radius (shown with the dashed circle), and are swept backward as they propagate outward. The magnetospheric material gradually mixes into the disk, and spirals sometimes merge as they evolve. The outer spiral in Figure~\ref{fig:centrifuge}(a) contributes to the large spiral in panel (b), while the strong pattern in panel (c) persists as the main spiral in panel (d), about three stellar spin periods later. 

The spiral patterns of magnetospheric material are visible in all variables sensitive to the relative strength of the magnetic field, including the density, plasma $\beta$, and the magnetization. However they are most clearly apparent in the passive scalar $\mathcal{F}$, which tracks gas that originated from the force-free magnetosphere where $\mathcal{F} = 1$ (Figure~\ref{fig:centrifuge}, bottom row). This is direct evidence that these low-density, high-magnetization structures are created by the intrusion of the magnetosphere into the accretion flow, rather than being formed in-situ by the disk's turbulence\footnote{Since the disk is much denser than the magnetosphere, only a small amount of mixed-in disk gas is necessary to significantly drop a cell's ``magnetospheric fraction'' $\mathcal{F}$. Spiral patterns are clearly visible down to the $\mathcal{F} \sim 10^{-7}$--$10^{-6}$ level.}.

This behavior should be more vigorous in the parallel orientation, because reconnection cannot easily relieve the accumulated magnetic pressure at the magnetospheric boundary. We generally see more signs of its presence, in the form of visible high-$\sigma$, high-$\mathcal{F}$ spirals from $\rmag$, in the parallel simulations. 

At $\mu=10$ there is significantly higher variability in the jet power in the parallel orientation (see Figure~\ref{fig:Ljet_time}), which may be because most of the additional jet magnetic flux must be opened by the intermittent interchange-slingshot process. 
As we saw in Figure~\ref{fig:torques}, the electromagnetic torque from field lines interacting with the accretion flow has a spin-up effect in the 10--antiparallel simulation, while it usually spins the star down in the 10--parallel run. In the antiparallel orientation, little stellar flux is transported deep into the disk, and so the accretion-interacting torque is dominated by the columns and the innermost part of the disk inside corotation. The parallel arrangement is more unstable to the slingshot behavior, and more stellar flux is dragged far into the disk beyond the corotation point, where it contributes a strong spin-down torque. 

There is generally more stellar magnetic flux coupling to the disk in the parallel orientation at small magnetic moment (Figure~\ref{fig:multivar_mu}), which may be understood as stellar field that has entered the outer layers of the disk by interchange motions. 

The strength of the interchange slingshot increases with increasing $\mu$ (or decreasing accretion rate). It is mostly inactive at $\mu=5$ for the reasons outlined above in case (a), but is able to open a moderately large fraction of the closed zone at $\mu=10$. By $\mu = 20$ it opens effectively all of the stellar field lines with which the disk comes into contact, causing the open flux and jet power to be insensitive to the field orientation. 

\begin{figure}
    \centering
    \includegraphics[width=\columnwidth]{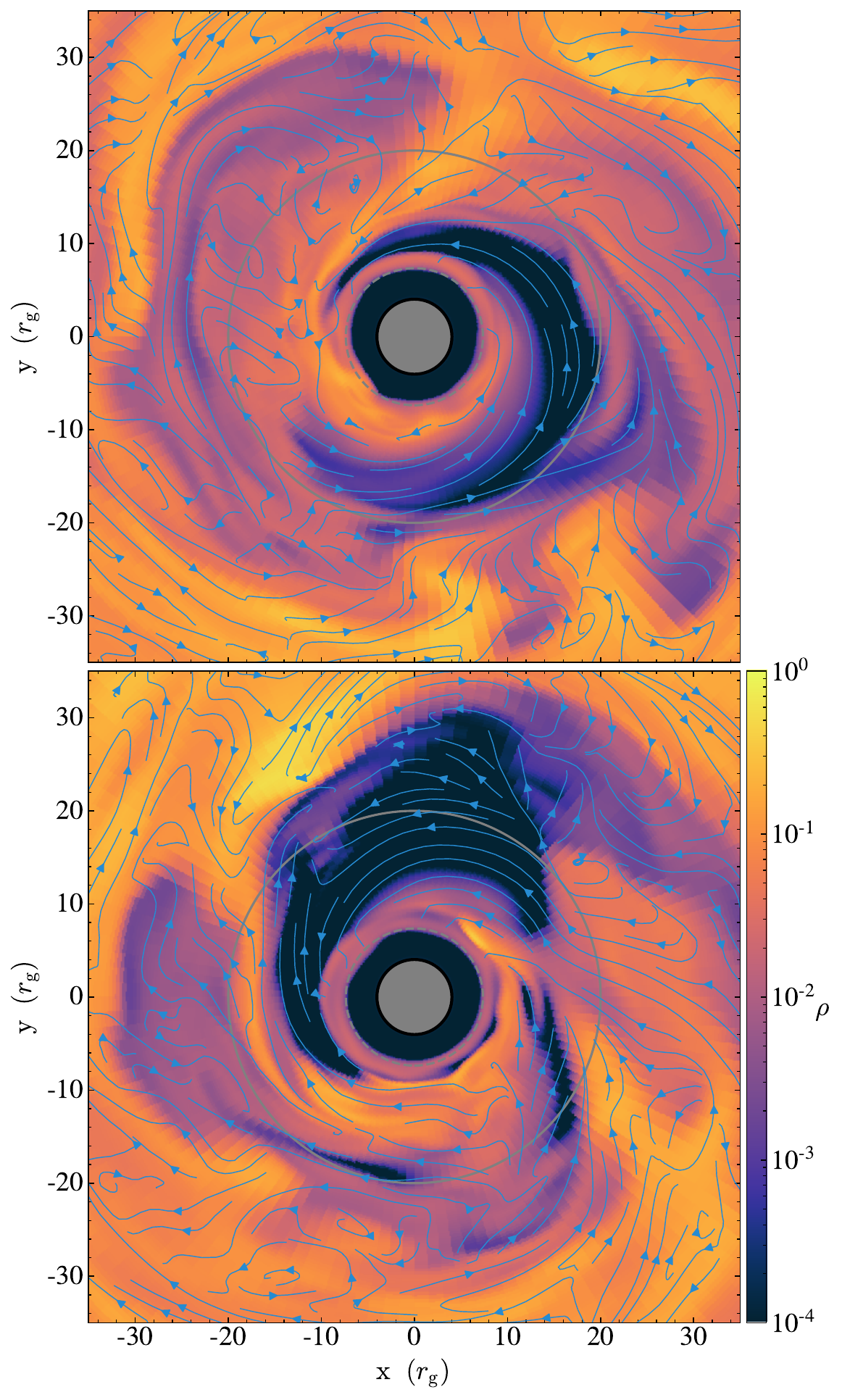}
    \caption{Large plumes of stellar magnetic field (low-density spiral regions) entering the accretion flow in the propeller-state 40--parallel simulation, shown in the equatorial plane. Top: $t = 17,\!060\rgc$; bottom: $17,\!280\rgc$.}
    \label{fig:mu40_slingshot}
\end{figure}

This is also true for the propeller-regime simulations at $\mu = 40$, where dramatic eruptions of magnetospheric field lines into the disk are observed in both orientations, as can be seen in Figure~\ref{fig:mu40_slingshot}. These large stellar-field plumes are likely to be an important contributor to matter ejection in the propeller state. 

In Figure~\ref{fig:10ar4_multivar_time}, the magnetospheric radius retreats as the large-scale accretion rate declines toward the end of the simulation. This coincides with an increase in the flux-opening efficiency $\zeta$ and hence in the jet power, and a decline in the disk-connected flux. This may be due to the interchange slingshot becoming more effective as $\rmag$ moves out toward $\rco$, allowing the disk-connected flux to open. Note that this is occurring in an antiparallel scenario.

\begin{figure}
    \centering
    \includegraphics[width=\columnwidth]{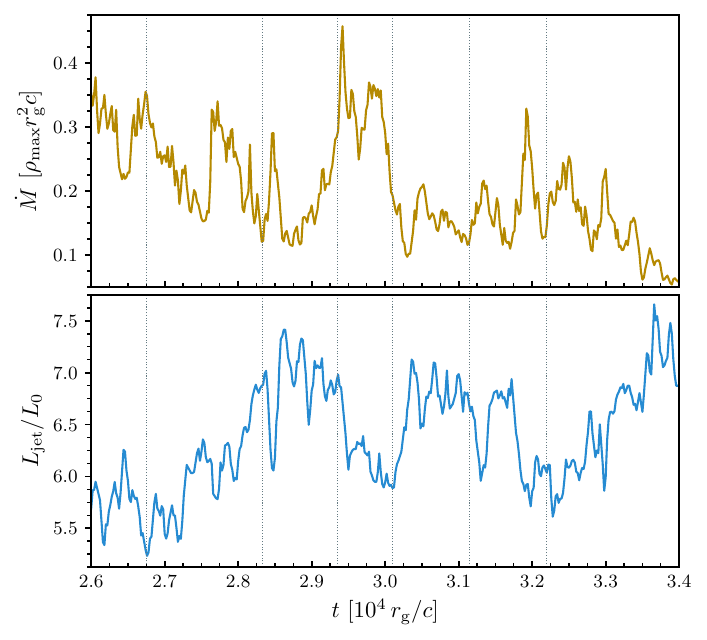}
    \caption{Anticorrelation between the stellar accretion rate and the jet power over short timescales in the 20--parallel simulation. Vertical dotted lines are to guide the eye to several examples of visible anticorrelation.}
    \label{fig:Lj_Mdot}
\end{figure}

The anticorrelation, over short timescales, between stellar accretion rate $\mdot$ and jet power is frequently observed; see Figure~\ref{fig:Lj_Mdot}. In particular, the 10--parallel and 20--parallel simulations have negative correlation coefficients between $\mdot$ and $\Lj$, measured roughly at the $-0.6$ level, over the period when the system is in an approximately steady state.

As an alternative to the process described above, one could imagine the disk's magnetic field, in the parallel star--disk orientation, twisting around in the $\phi$ direction until it is antiparallel to the magnetospheric field lines and can open by reconnection. However we see no evidence of large-scale changes in the direction of the disk's poloidal field. The snapshots in Figure~\ref{fig:mu5-40_poloidal} (bottom row) show mostly poloidal field in the same direction as in the initial conditions, as is also observed in the azimuthal and temporal averages.

\subsection{Accretion flow properties}
\label{sec:afprops}

\begin{figure}
    \centering
    \includegraphics[width=\columnwidth]{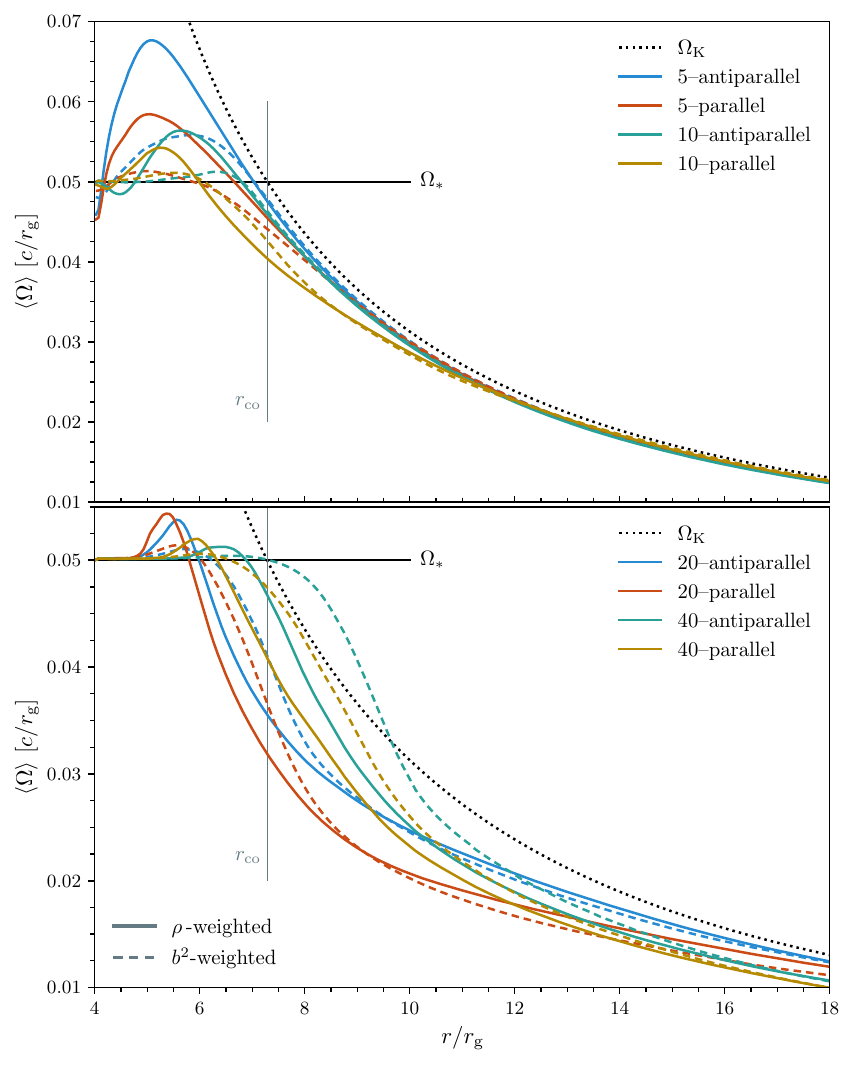}
    \caption{Average angular velocity profiles at the equator. Solid lines represent density-weighted averages, dashed lines are weighted by $b^2$. The solid horizontal lines indicates the stellar spin $\Omegastar$, and the dotted black lines the Keplerian angular velocity.}
    \label{fig:omega}
\end{figure}

We have primarily focused on how the presence of the disk affects the star, its magnetosphere, and the surrounding region. However the star-disk interaction can also lead to an accretion flow that is substantially different to what would be found around an unmagnetized accretor. In our simulations this occurs predominantly by the transport of the star's magnetic field into the disk.

In Figure~\ref{fig:omega} we plot the radial profile of the gas angular velocity within $\Delta \theta = \pm \pi/24$ of the equator, averaged azimuthally and over the steady-state period $\dtsteady$. The solid lines are weighted by the matter density, while the dashed lines are weighted by the fluid-frame magnetic energy density, $b^2/2$. The nominal corotation radius $\rco$ is where the Keplerian rotation rate $\Omegak$ equals the stellar angular velocity $\Omegastar$, as indicated by the thin vertical line. 

The $\rho$- and $b^2$-weighted angular velocities are generally nearly equal at the point where they match the stellar spin angular velocity. This is the ``true'' corotation point, which in these simulations lies inside $\rco$ because these partially pressure-supported disks are sub-Keplerian. Inside this point any stellar magnetic field that is coupled to the disk will act to slow the gas, and causes $\Omegab < \Omegarho$. Outside this point the star-coupling field pulls the gas forward, giving $\Omegab > \Omegarho$. 

The mismatch $\Omegab - \Omegarho$ first increases with increasing $r$, as the stellar field tries to pull the material forward at a constant $\Omegastar$ while the natural rotation profile of the disk declines. Eventually most star-disk coupling field lines are opened by the velocity shear, and the star ceases to apply a torque to the disk; this causes the angular velocity mismatch to decrease. Eventually a point is reached where the two averages are equal, and beyond which $\Omegarho > \Omegab$ again; this may be because magnetic braking is stronger in the higher-magnetic-field regions. 

This pattern is visible to a greater or lesser degree in every simulation (Figure~\ref{fig:omega}); the two $\mu=20$ runs being particularly clear examples. The difference between the two averages is larger in the $\mu = 20$ and 40 runs, which is consistent with there being more stellar flux interacting with the disk, as argued above. 

The two $\mu = 5$ simulations, and 10--antiparallel, are quite close to the Keplerian profile. The 10--parallel run is more sub-Keplerian, possibly due to there being more pressure support from additional intruding stellar magnetic field since the interchange slingshot is more active. The two $\mu=20$ simulations have accretion flows that are even more sub-Keplerian, again due to enhanced pressure support from the stellar field. The trends reverses by the propeller-state $\mu=40$ scenarios, where azimuthal acceleration from the star-disk coupling field appears to overcome the slowing effects of additional pressure support. In particular, the magnetic-energy-weighted angular velocity in the 40--antiparallel simulation is the only profile that shows \emph{super-Keplerian} rotation. One may expect super-Keplerian rotation to be the norm, at least in the inner disk, in strong propeller cases where $\rmag \gg \rco$.

\begin{figure}
    \centering
    \includegraphics[width=\columnwidth]{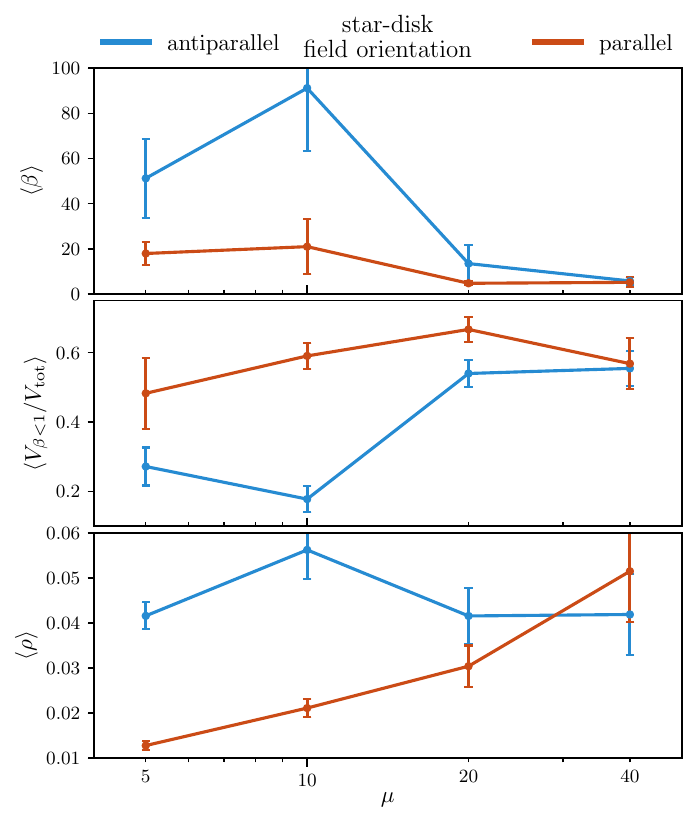}
    \caption{Average disk properties inside the light cylinder, where the disk is defined as the $\sigma<1$ region, as a function of stellar magnetic moment. From top: plasma $\beta$; the fraction of the total included disk volume in which $\beta < 1$; and the mass density. Vertical bars indicate standard deviations over the averaging period.}
    \label{fig:disc_properties}
\end{figure}

Figure~\ref{fig:disc_properties} shows average properties of the inner accretion flow, as a function of stellar magnetic moment. The accretion flow is defined as the $\sigma < 1$ region, and the average includes material inside the light cylinder, $r\,\sin\theta < \RLC$, taken over the steady-state period $\dtsteady$. Again, we see a clear dependence on the initial star-disk field orientation at smaller magnetic moment, and much less dependence at higher $\mu$. 

At $\mu = 20$ and 40 we find that the average plasma $\beta$ is low, $\langle\beta\rangle \sim 5$. At $\mu = 5$ and 10, it is also comparatively low, $\sim 20$, in the parallel orientation, while it reaches much higher values in the antiparallel simulations. Similarly, at low $\mu$ the average fraction of the disk volume in which $\beta < 1$ is  higher in the parallel orientation, $\langle V_{\beta<1}/V_{\rm tot}\rangle \sim 0.5$--0.6, than in the antiparallel case. The fraction is universally high at larger $\mu$. The mass density higher by a factor of 3--4 in the antiparallel case at low $\mu$, while the difference disappears at larger magnetic moment.

Again, this can be understood as a response to the various configurations' susceptibility to star-disk magnetic coupling and the interchange slingshot, as functions of $\rmag/\rco$ and the relative field orientation. The 5--antiparallel and 10--antiparallel runs are only weakly unstable to interchange and the slingshot, since $\rmag \lesssim \rco$ and field lines can relieve pressure at the magnetospheric boundary by reconnecting. This results in dense, high-$\beta$ disks. 

The 5--parallel and 10--parallel simulations are much more susceptible to stellar field invading the disk through interchange, reducing the average plasma $\beta$. In the 5--parallel case $\rmag \ll \rco$ and the slingshot is inactive, and so the stellar field only mixes into the innermost part of the disk, puffing it up (Figure~\ref{fig:mu5-40_poloidal}, bottom-left panel) and dropping the mass density. The interchange slingshot occurs more easily in the 10--parallel simulation, and so while $\beta$ remains low the disk does not inflate as much, and the density is higher. At $\mu = 20$ and 40 both orientations are unstable to the slingshot, giving low-$\beta$ inner disks in which a large volumetric fraction is dominated by magnetic pressure.

We suggest that these dependences, on $\mu$ and orientation, of the angular velocity profiles and average disk quantities constitute additional circumstantial evidence for the movement of stellar magnetic flux through the disk via a mechanism like that described in Section~\ref{sec:interchange}.

\section{Discussion and Conclusions}
\label{sec:conclusions}

We have presented the first 3D general-relativistic MHD simulations of accretion onto rotating neutron stars, using a method allowing highly magnetically dominated, nearly force-free, regions to evolve naturally alongside dense MHD flows. These capture several important effects that were missing in previous axisymmetric studies \citep[\hyperlink{PT17target}{PT17};][]{2022MNRAS.515.3144D}, such as the generation of self-sustaining magnetorotational turbulence in the accretion flow, the ability of the star's closed magnetosphere to penetrate the disk through non-axisymmetric interchange modes, and the freedom of the disk's magnetic field to spontaneously change its orientation with respect to the stellar field. They also differ from existing non-relativistic 3D simulations \citep{2012MNRAS.421...63R,2021MNRAS.506..372R} in conserving total energy rather than entropy, and therefore having a thick accretion flow, and in the exclusion of the star's magnetic field from the initial matter distribution. This field arrangement, and the placement of all of the accretable mass outside the star's light cylinder in the initial conditions, permits a direct test of the importance of the relative polarity of the star's dipole and the disk's poloidal field. This effect is large in axisymmetry \citep{2011MNRAS.416..416R}, with completely different magnetic configurations being produced depending on whether the stellar and disk fields met in parallel or antiparallel orientations (\hyperlink{PT17target}{PT17}). 

We find that the relative-field-orientation effect is more complicated in 3D. Generally, it persists when the stellar field is weak compared to the accretion rate, or in other words when the magnetospheric boundary lies well within the corotation radius. In this arrangement, when the fields are mutually antiparallel the stellar closed field lines are opened by reconnection and a strong relativistic jet is launched. When they are parallel the stellar field closes, and no jet is produced, as in the axisymmetric simulations. However when the star's magnetic dipole moment is larger, or the accretion rate is lower, the magnetospheric boundary can be near or beyond the corotation point and a different state is found. Now the stellar field is opened efficiently by its interaction with the disk and a jet is produced, even in the parallel orientation, and most physical quantities become independent of the star-disk relative field polarity. 

The opening of the star's dipolar field lines by the disk, when they are not able to reconnect, may be due to an effect we have termed the interchange slingshot. Stellar field lines can enter the disk at the magnetospheric radius $\rmag$ through non-axisymmetric modes of the interchange instability. When $\rmag$ is deep inside the corotation point the star-disk magnetic coupling causes the nearby disk gas to lose angular momentum and spiral inward, destroying the perturbation. When $\rmag$ is beyond corotation the coupling pushes the disk gas to spiral outward, clearing the way for the stellar-field perturbation, which is also driven outward by stresses due to the pinched shaped of the inflating field lines, to grow into an expanding plume. The stellar field is transported deep into the disk, and is rapidly opened by twisting due to the angular velocity mismatch between star and disk. This interpretation appears to be consistent with how various system properties --- such as the amount of open stellar flux, the torque applied by field lines coupling the star and disk, the angular velocity profile of disk material, and the balance between thermal and magnetic pressure in the disk --- depend on the relative field polarity and the location of the magnetospheric radius, as it varies due to the stellar field strength or the instantaneous large-scale accretion rate in a simulation.

The spiral plumes of magnetospheric magnetic field visible in Figures~\ref{fig:centrifuge} and \ref{fig:mu40_slingshot} appear to be similar to flux eruption events in black-hole accretion flows in the magnetically arrested disk state \citep{2008ApJ...677..317I, 2011MNRAS.418L..79T,2022ApJ...924L..32R,2022ApJ...941...30C}. The behavior here may be somewhat different, as the magnetic field in the black hole's magnetosphere is primarily vertical rather than dipolar, and more importantly is not frozen into a rotating surface. The interchange slingshot mechanism, as outlined in Section~\ref{sec:interchange}, would not operate in its proposed form in a black-hole system, as it relies on the communication between field lines' footpoints on the star and in the disk to produce distinct outcomes at high ($\rmag < \rco$) and low ($\rmag > \rco$) accretion rates. In the future, it would be interesting to compare black-hole and neutron-star accretion to investigate the effect of the line-tied boundary condition. 

The flux-opening efficiency factor $\zeta$ is comparatively high, $\zeta \gtrsim 0.5$--0.75, in most of the scenarios we simulated (Fig.~\ref{fig:multivar_mu}). It is generally higher, approaching unity, when the system is in the propeller regime, as was previously found in idealized prescribed-disk simulations by \citet{2017MNRAS.469.3656P}. For the antiparallel star--disk orientation these high $\zeta$ values are roughly consistent with the axisymmetric results of \hyperlink{PT17target}{PT17}, and are higher than the range $\zeta \sim 0.3$--0.5 obtained by \citet{2022MNRAS.515.3144D}; the latter study may have observed less flux opening because the initial dipole field threaded the torus rather than being diverted around it as in \hyperlink{PT17target}{PT17} and the present work. 

In the parallel orientation, however, the outcomes are in some cases very different to the axisymmetric results. For scenarios in which the accretion flow is truncated close to or beyond the corotation radius we recover nearly identical $\zeta$ values to the antiparallel case, indicating strong flux opening, while in \hyperlink{PT17target}{PT17} flux closing was seen in all parallel-orientation simulations. We ascribe this 3D-only flux opening to non-axisymmetric interactions between the magnetosphere and disk leading to the interchange slingshot mechanism. 

The only circumstance in which we do not obtain robust flux opening is when the initial torus field is parallel to the dipole, and the effective accretion rate is high enough to push the disk to well within the corotation radius. In this case flux closing is observed, as in axisymmetry. The strong dependence of the flux-opening efficiency, and hence the jet power, on the magnetic orientation in the high-accretion-rate regime raises the possibility of significant jet variability in these systems, as one would expect that over time the accretion flow would drag in flux systems with varying magnetic polarity.  

Similarly, for accreting systems one might also expect there to be observable differences in the disk emission due to the relative magnetic orientation, since it can have an effect on the disk's average bulk properties such as density and plasma $\beta$ (Figure~\ref{fig:disc_properties}). 

Our $\mu = 20$ simulations are in a state intermediate between accreting and propeller regimes, while at $\mu = 40$ they are clearly in the propeller state for most of their duration. While there have been numerous axisymmetric studies of the propeller regime with non-relativistic \citep{2004ApJ...616L.151R,2005ApJ...635L.165R,2018NewA...62...94R,2006ApJ...646..304U,2014MNRAS.441...86L} and relativistic \citep[\hyperlink{PT17target}{PT17};][]{2022MNRAS.515.3144D} simulations, these are the first 3D simulations to explore this state. We find that relativistic jets are launched in the propeller regime, just as in the accreting state, by the star's rotation and stellar flux opened by the star--disk interaction \citep{2016ApJ...822...33P}. The efficiency of flux opening is very high in the propeller regime, with $\zeta$ values approaching unity. There is little dependence on the star--disk relative field orientation for most of the system's average properties ($\dot{M}$, $\Lj$, $N$, $\zeta$, $\langle\beta\rangle$, $V_{\beta<1}$, $\rho$), because of the vigorous operation of the interchange slingshot when the system starts in the parallel orientation. We measure more variability in $\rmag$ in the antiparallel case (Figure~\ref{fig:multivar_mu}), which may be related to flux opening being driven by two competing mechanisms (direct reconnection and interchange).

The propeller simulations generally showed stronger non-axisymmetric features than the accreting state, with the most pronounced structures occurring in the parallel orientation (Figure~\ref{fig:mu40_slingshot}). We also ran exploratory simulations at higher stellar field strengths, $\mu = 80$ and 160, that approximately reproduced the strong propeller and ``radio ejection'' (exclusion of the accretion flow from the light cylinder) states seen in our axisymmetric study. There was, again, greater disruption of the disk by non-axisymmetric incursions by the stellar magnetosphere than at lower stellar magnetic moment. We defer a detailed investigation of the behavior at low effective accretion rate to future work.

Our results have been limited to the special case in which the star's spin and magnetic axes are aligned. We would expect the conclusions to transfer with only minor quantitative adjustments to systems with small spin--magnetic obliquity angles $\lesssim 15^\circ$, beyond the breaking of the quasi-axisymmetric accretion ``curtains'' around the poles into two accretion streams with limited extent in the $\phi$ direction \citep{2003ApJ...588..400R}. At larger obliquity more significant deviations can be expected, since as obliquity increases more of the star's closed magnetosphere ceases to interact with the accretion flow. We defer an investigation of accretion onto oblique pulsars to a subsequent work.

%%%\begin{acknowledgments}
\vspace{5mm}
This work was supported by the NASA Astrophysics Theory Program, grant no.\ 80NSSC21K1746. The simulations were performed on the Pleiades cluster, provided through the NASA Advanced Supercomputing (NAS) Division, under project s5140, and on the Stampede2 cluster at the Texas Advanced Computing Center (TACC) under allocation AST150062. The research described in this paper was conducted under the Laboratory Directed Research and Development Program at Princeton Plasma Physics Laboratory, a national laboratory operated by Princeton University for the U.S.\ Department of Energy under Prime Contract No.\ DE-AC02-09CH11466. We acknowledge support by the NSF through resources provided by TACC Stampede2, where simulations were also
carried out, and Frontera \citep{stanzione2020frontera}, where data were analyzed, and NCSA
MSS and TACC Ranch, where data were backed up, under grants TG-AST100040 (TeraGrid), AST20011 (LRAC), and AST22011 (Pathways).
%%%\end{acknowledgments}

\hypertarget{PT17target}{}
\bibliography{main}{}
\bibliographystyle{aasjournal}

\appendix

\section{Deforming the Stellar Dipole}
\label{sec:deform}

In the initial conditions, the dipole field is deformed around the torus by setting the vector potential inside the torus to a constant value $A_{\rm closed}$, approximately equal to the vector potential $A_\phi(r,\theta) = 2\pi \int_0^\theta B^r(r,\theta')\,\sqrt{\gamma}\, {\mathrm d}\theta'$ at $(r,\theta) = (\RLC, \pi/2)$ in the steady state of the equivalent isolated pulsar (i.e.\ one having the same $\Omega$ and $\mu$ but without an accretion flow). A thin equatorial channel connecting the torus to the inner magnetosphere also has its vector potential set to $A_{\rm closed}$. We can manipulate the magnetic flux surfaces using a single component of the vector potential because our initial conditions are axisymmetric.

\begin{figure}
    \centering
    \includegraphics[width=\columnwidth]{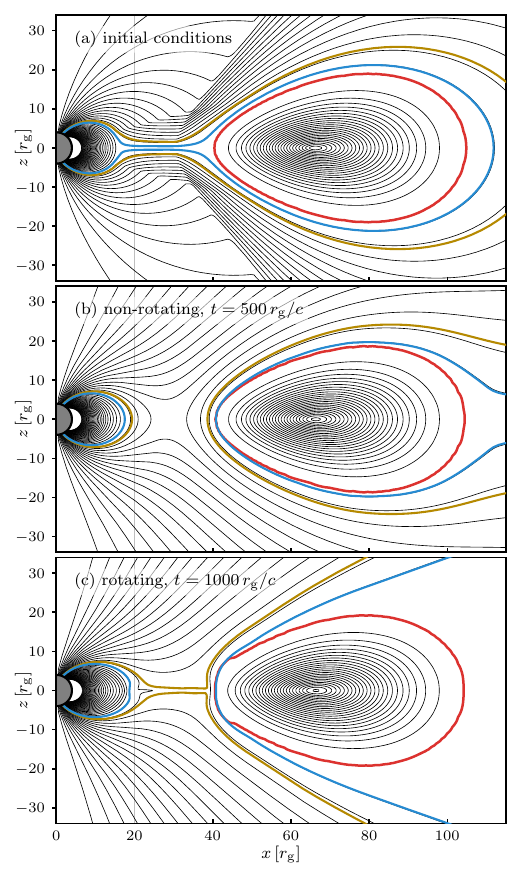}
    \caption{Deformation of the stellar dipole field around the torus in the 10--antiparallel simulation. (a) Field lines that are open in the isolated-pulsar state are wrapped around the far side of the torus in the initial conditions. (b) The stellar field relaxes, without rotation, for the first 500 $\rgc$ of the simulation. (c) Once rotating, the magnetosphere quickly takes the form of the isolated pulsar near and inside the light cylinder (vertical gray line), with open field lines being deflected around the torus. Black lines represent azimuthally averaged poloidal magnetic flux surfaces, equally spaced in $\langle A_\phi \rangle$; colored lines show the approximate last open flux surfaces in the isolated state (blue) and with the torus (yellow). The torus is indicated by a density contour, in red, at $\rho = 10^{-3}\, \rho_{\rm max}$. }
    \label{fig:deform}
\end{figure}

The shape of the torus is supplied by an approximate effective potential $\Phi$, modeled on that of a Newtonian vorticity-free torus,
\begin{equation}
    \Phi(R, Z) = - \frac{G \MNS}{\sqrt{a^2 R^2 + b^2 Z^2}} + \frac{l^2}{2 R^2},
\end{equation}
where $(R,Z)$ are cylindrical coordinates and the constants $a$ and $b$ are chosen depending on the torus's shape; for our torus we use $a = 0.975$ and $b = 1.85$. The specific angular momentum is set by the pressure-maximum radius, $l^2 = G\MNS\rmax$, since orbits are Keplerian at $(R,Z) = (\rmax,0)$. We associate the torus surface with the effective potential at the inner edge, $\Phi_{\rm in} = \Phi(\rinner,0)$, and define a second value to limit the volume in which the field is deformed, $\Phi_{\rm out} = 0.2 \Phi_{\rm in}$.

The deformation factor $f(R,Z)$ is first determined by the value of $\Phi = \Phi(R,Z)$ at each point,
\begin{equation}
    f = 
    \begin{cases}
        0 & \textrm{if } \Phi < \Phi_{\rm in} \\ 
        \frac{\Phi - \Phi_{\rm in}}{\Phi_{\rm out} - \Phi_{\rm in}} & \textrm{if } \Phi_{\rm in} < \Phi < \Phi_{\rm out} \\
        1 & \textrm{otherwise}.
    \end{cases}
\end{equation}
The equatorial channel is then constructed, using a modification width $\delta_\theta = \pi/12$; the channel is applied to that part of the region $r < \rmax$ in which $A_\phi < A_{\rm closed}$. The cells immediately adjoining the equator are set to $f = 0$, while those within $\delta_\theta$ pick up an additional multiplicative term, 
\begin{equation}
    f \leftarrow f \times \frac{|\theta - \pi/2|}{\delta_\theta} \quad \textrm{if} \quad |\theta - \pi/2| < \delta_\theta .
\end{equation}
The new, deformed-field vector potential $\tilde{A}_\phi$ at every point is then found from its original value $A_\phi$ using
\begin{equation}
    \tilde{A}_\phi = f\, A_\phi + \left(1 - f\right)\, A_{\rm closed}.
\end{equation}

The results of this procedure are shown in Figure~\ref{fig:deform}(a). For the simulations' first 500 $\rgc$ the star does not rotate and the magnetosphere relaxes, producing the smooth and nearly potential configuration in Figure~\ref{fig:deform}(b). At this time stellar rotation begins, rising linearly to its final value over 40 $\rgc$. The magnetosphere quickly reaches a steady state which for $R \ll \rinner$ is very similar to that of an isolated rotating neutron star \citep{1999ApJ...511..351C,2006MNRAS.367...19K,2006ApJ...648L..51S}, with the open field lines being deflected around the torus as in Figure~\ref{fig:deform}(c). There is no rotationally opened flux trapped inside $\rinner$, because the open flux in the final state is slightly lower than that originally wrapped around the torus.

\section{Hybrid MHD--Force-free Approach}
\label{sec:app_method}

Standard MHD evolution using conservative equations becomes unstable when the energy density in the magnetic field becomes superdominant, and small errors in the magnetic field lead to large errors in the hydrodynamic quantities; this generally occurs when $\sigma \gtrsim 100$. The energy density in the dipole field of a neutron star scales as $r^{-6}$, so a dynamically important stellar field at large radii (e.g., at the light cylinder) implies a very strong field near the surface, which would require a dense stellar atmosphere if the magnetization were to be kept low enough to maintain stability.

An alternative is to use force-free electrodynamics (FFE), the $\sigma \rightarrow \infty$ limit of plasma dynamics \citep{2002MNRAS.336..759K}, in the magnetosphere. In this system the hydrodynamic quantities are negligible, and the only variables are the magnetic field and either the inductive electric field or, equivalently, the velocity of magnetic field lines. In other words, the only velocity is that perpendicular to the magnetic field --- there is no concept of velocity parallel to $B$. 

Rather than evolving the MHD and FFE systems in separate regions of the domain we instead evolve the MHD system (Equations~\ref{eq:mhd}) everywhere, and at the end of each substep adjust the solution in the magnetosphere to damp the degrees of freedom not present in FFE.

The degree of adjustment is determined by two factors: the values of the passive scalar $\mathcal{F}$ and the fixed radial profile $\mathcal{P}(r)$. The ``magnetospheric fraction'' $\mathcal{F}$ is initially set to 0 inside the torus and to 1 in the surrounding magnetosphere. It represents the fraction of the density in a cell that is ascribed to the force-free magnetosphere, and that therefore can be increased or decreased without affecting the conservation of the physically modeled matter density in the accretion flow.

We aim to minimize the departures from evolution with the full MHD equations, and therefore restrict the FFE-like behavior to a sphere of radius equal to the star's light cylinder. The radial profile
\begin{equation}
    \mathcal{P}(r) = 
    \begin{cases}
        1 - \sin^4\left( \frac{\pi}{2} \frac{r-\rNS}{\rlc-\rNS}\right) & \textrm{if } r \leq \rlc \\
        0 & \textrm{if } r > \rlc
    \end{cases}
\end{equation}
has $\mathcal{P}(\rNS) = 1$ and is zero at and beyond the ``light sphere.'' We make all adjustments using the combination
\begin{equation}
    \mathcal{A} = 1 - \mathcal{F}\mathcal{P}.
\end{equation} 
When $\mathcal{A} = 1$ the flow is unmodified, which occurs either inside the accretion flow ($\mathcal{F}\approx 0$) or in any cell outside the light sphere ($\mathcal{P} = 0$). $\mathcal{A} = 0$ gives maximal suppression of non-FFE degrees of freedom, and $0 < \mathcal{A} < 1$ interpolates smoothly between MHD and FFE behavior. 

We set background distributions of density and internal energy forming an approximate hydrostatic atmosphere near the star and falling off like $r^{-6}$ at large radii, with a smooth transition around $r = \rlc$. These distributions, $\rhobg$ and $\ubg$, function as the simulation's floors. The normalization scales with $\mu^2$ such as that the magnetization $\sigma \gtrsim 10^4$ near the star.

\subsection{Density and internal energy adjustment}

At the end of each substep, over which the MHD equations are evolved as usual, we perform a combined flooring and force-free-ification procedure. The pre-adjustment density $\rho_0$ and internal energy $\epsilon_0$ values are stored temporarily. The density is separated into its ``magnetospheric'' and ``accretion flow'' components,
\begin{align}
    \rhom &= \rho \, \mathcal{F} ,\\
    \rhoaf &= \rho \, (1 - \mathcal{F}) ,
\end{align} 
with the intention that only the magnetospheric component $\rhom$ is ever modified. 

If $\rho < \rhobg$, the density is increased to $\rhobg$ and the difference is allotted to $\rhom$: 
\begin{align}
\rho &\leftarrow \rhobg\\
\rhom &\leftarrow \rhobg - \rhoaf. \label{eq:rhom_floor}
\end{align}

If $\rhom > \rhobg$ the amount of magnetospheric gas can be reduced,
\begin{equation}
    \rhom \leftarrow \rhobg + (1 - \mathcal{P}) (\rhom - \rhobg),\label{eq:rhom_drain}
\end{equation}
with a resulting reduction in total density, 
\begin{equation}
    \rho \leftarrow \rhom + \rhoaf.
\end{equation}
Unlike the previous operation, this adjustment is subject to the radial profile $\mathcal{P}$ and so isn't applied outside the light sphere.

A floor is also applied to the internal energy: if $\epsilon < \ubg$ replace 
\begin{equation}
    \epsilon \leftarrow \ubg.
\end{equation}
If $\epsilon > \ubg$, magnetospheric gas can be cooled toward the floor: 
\begin{equation}
    \epsilon \leftarrow \ubg + \mathcal{A} (\epsilon - \ubg).
\end{equation}
The use of $\mathcal{A}$ here restricts the cooling to cells that are largely magnetospheric ($\mathcal{F} \sim 1$) and near the star.
The new values of density and internal energy, $\rho_1$ and $\epsilon_1$, can now be stored in the main arrays.

\subsection{Velocity adjustment}
\label{sec:velocity_adjustment}

The fluid velocity along the magnetic field is modified for two purposes: (a) if $\rho_1 > \rho_0$ or $\epsilon_1 > \epsilon_0$ we reduce the parallel velocity so that the conserved momentum along the magnetic field is unchanged; (b) we reduce the parallel velocity of magnetospheric gas inside the light sphere to suppress this non-force-free degree of freedom and improve stability at high magnetization. We apply this procedure in the frame of the observer $\hat{\eta}^\mu$ that is static with respect to the coordinates,
\begin{equation}
    \hat{\eta}^{\mu} = \left( \frac{1}{\sqrt{-g_{tt}}}, 0^i \right).
    \label{eq:coordstaticobs}
\end{equation}
Here and in the rest of the appendices we set $c=1$.
The magnetic field measured by this observer is $\hat{b}^\mu = - \hat{\eta}_\nu \tensor[^*]{F}{^{\mu \nu}}$. Using the dual of the electromagnetic field tensor expressed in terms of the fluid velocity and fluid-frame magnetic field, $u^\mu$ and $b^\mu$,
\begin{equation}
    \tensor[^*]{F}{^{\mu \nu}} = b^\mu u^\nu - b^\nu u^\mu,
\end{equation}
one can construct the static-observer-measured magnetic field as
\begin{equation}
    \hat{b}^\mu = \left(b\cdot \hat{\eta}\right)\, u^\mu - \left(u\cdot \hat{\eta}\right)\, b^\mu,
    \label{eq:bstatic}
\end{equation}
using the notation $x\cdot y = x^\alpha y_\alpha$.
We define the components of velocity parallel and perpendicular to the magnetic field as
\begin{align}
    u_\parallel^\mu &= \frac{u \cdot \hat{b}}{\hat{b}^2} \, \hat{b}^\mu,\\
    u_\perp^\mu &= \left( \delta^\mu_\beta - \frac{\hat{b}^\mu \hat{b}_\beta}{\hat{b}^2}\right)\, u^\beta
\end{align}
where $\delta^\mu_\nu$ is the identity tensor; $u^\mu = u_\parallel^\mu + u_\perp^\mu$ by construction\footnote{If the static observer is replaced by the hypersurface-normal fiducial observer, $\hat{\eta}_\mu \rightarrow n_\mu$, the perpendicular velocity component constructed here is, when normalized, identical to the drift-frame velocity of \citet{2009ApJ...697.1164B}}. We wish to preserve $u_\perp^\mu$ and reduce $u_\parallel^\mu$, and so will be replacing the fluid velocity with the updated vector
\begin{equation}
    \tilde{u}^\mu = \Gamma \left( u_\perp + \lambda u_\parallel^\mu\right),
    \label{eq:tildeu}
\end{equation}
where $\lambda \leq 1$ and $\Gamma = 1/\sqrt{-\left(u_\perp^2 + \lambda^2 u_\parallel^2\right)}$ is chosen to normalize $\tilde{u}^2 = -1$.

Our first aim is to preserve the conserved momentum along $\hat{b}^\mu$,
\begin{equation}
    \hat{b}^\nu \tensor{\mathcal{T}}{^{\,t}_{\,\nu}} = K,
    \label{eq:consmom}
\end{equation}
following the addition of mass or internal energy due to the floors; $\mathcal{T}^{\alpha\beta} = (\rho + \epsilon + p)\, u^\alpha u^\beta + p\, \delta^{\alpha\beta}$ is the hydrodynamic energy-momentum tensor. 

The value of $K$ is set using the pre-flooring values of density and internal energy, $\rho_0$ and $\epsilon_0$, and the original fluid velocity $u^\mu$. Then the adjusted energy-momentum tensor $\tilde{\mathcal{T}}^{\alpha\beta}$ is created using $\rho_1$, $\epsilon_1$, $p_1 = (\gamma -1)\epsilon_1$, and $\tilde{u}^\alpha$; inserting this into equation~(\ref{eq:consmom}) and expanding $\tilde{u}^\mu$ with equation~(\ref{eq:tildeu}) gives a quadratic equation for $\lambda$,
\begin{equation}
    \left(u_\parallel^t + \kappa u_\parallel^2\right)\,\lambda^2 + u_\perp^t\,\lambda + \kappa u_\perp^2 = 0,
    \label{eq:quadratic}
\end{equation}
where
\begin{equation}
    \kappa = \frac{K - p_1 \hat{b}^t}{\left(\hat{b}\cdot u_\parallel\right)\, h_1}
\end{equation}
and $h_1 = \rho_1 + \epsilon_1 + p_1$ is the post-floor hydrodynamic enthalpy density. Equation~(\ref{eq:quadratic}) can be solved with the standard formula, taking the root lying in the range $\lambda = [0,1]$.

Once $\lambda$ has been found the second adjustment, to reduce the parallel velocity in the force-free magnetosphere, can be performed. This is as simple as reducing the parallel component by a factor of $\mathcal{A}$,
\begin{equation}
    \lambda \leftarrow \mathcal{A}\,\lambda.
\end{equation}
One can now construct the new fluid 4-velocity $\tilde{u}^\mu$ using $\lambda$ and the unchanged $u_\perp^\mu$ in equation~(\ref{eq:tildeu}), and set $u^\mu \leftarrow \tilde{u}^\mu$.

Finally, the magnetospheric fraction can be updated using the new value of $\rhom$ from equations~(\ref{eq:rhom_floor}) or (\ref{eq:rhom_drain}),
\begin{equation}
    \mathcal{F} \leftarrow \frac{\rhom}{\rhom + \rhoaf}.
\end{equation}

\section{Boundary Conditions}
\label{sec:app_BCs}

The \textsc{harmpi} code evolves the primitive variables at cell centers using fluxes calculated at cell interfaces. The neutron-star surface boundary conditions are applied by setting the primitive variables on the interface forming the inner edge of the domain, half a cell below the first cell's center. 

For a given radial line of cells, having centers at the same $\theta$ and $\phi$ coordinates, we label the first cell of a generic primitive variable $p$ as $p_0$, with the two succeeding cells moving into the domain being $p_1$ and $p_2$; $p_{\rm surf}$ is the value at the interface coincident with the stellar surface. For the purposes of the Lax-Friedrichs flux we set the left and right states to the same value, $p_{\rm l} = p_{\rm r} = p_{\rm surf}$.

Two methods are used for setting the primitive variables on the surface. If the mass density in the first cell is largely magnetospheric ($\mathcal{F} > 0.5$) or the radial velocity is directed outwards ($u^r > 0$) the ``force-free'' boundary condition prescribes a background hydrostatic atmosphere and enforces rotation at the stellar angular velocity $\Omegastar$. Otherwise we use an ``accreting'' boundary condition. 

\subsection{Densities and magnetic field}

Several variables are extrapolated to the boundary using the slope $\Dp$:
\begin{equation}
    p_{\rm surf} = p_0 - \frac{\Dp}{2}.
    \label{eq:prim_extrap}
\end{equation}

If a surface cell is accreting, $\rho$ and $\epsilon$ use slope-limited extrapolation with the monotonized-central limiter ($\mathrm{MCL}$),
\begin{equation}
    \Dp = \mathrm{MCL}(p_0,p_1,p_2).
    \label{eq:slopelim}
\end{equation}

If the cell is using the force-free boundary condition, $\rho$ and $\epsilon$ on the boundary interface are set using the same approximate hydrostatic atmosphere as forms the background for the rest of the domain. 

The magnetic field components $B^\theta$ and $B^\phi$ are extrapolated with the simple slope
\begin{equation}
\Dp = p_1 - p_0,
\end{equation}
while the radial magnetic field at the surface is set from the distribution stored at the beginning of the simulation using the values in the first shell of cells, $B^r_0(\theta)$,
\begin{equation}
    B^r_{\rm surf}(\theta) = B^r_0(\theta) \, \left(\frac{r_0}{\rNS}\right)^3
\end{equation}
where $r_0$ is the radial coordinate of the first cell's center.

\subsection{Velocity field}

In the accreting boundary condition, we go into the frame of the rotating stellar surface given by
\begin{equation}
    \us^\mu = (\us^t, 0, 0, \Omegastar \us^t)
\end{equation}
where normalization to $\us\cdot\us=-1$ gives
\begin{equation}
    \us^t = \frac{1}{\sqrt{- (g_{tt} + 2 g_{t\phi}\Omegastar + g_{\phi\phi} \Omegastar)}}.
\end{equation}
In this frame we extrapolate the fluid velocity parallel to the magnetic field, $u \cdot \bs$, to the stellar surface. 

The surface-observer magnetic field $\bs^\alpha$ is found by recognizing that the electric field is zero in the frame of the perfectly conducting surface, and so the dual electromagnetic field tensor can be represented as
\begin{equation}
    \tensor[^*]{F}{^{\mu \nu}} = \bs^\mu \us^\nu - \bs^\nu \us^\mu.
\end{equation}
Contracting with the hypersurface-normal observer $n_\nu$ gives
\begin{equation}
    -b_{\rm n}^\mu = (\us\cdot n) \,\bs^\mu - (\bs\cdot n)\, \us^\mu,
\end{equation}
where $b_{\rm n}^\mu = (0, B^i)$ is the normal-observer-measured field. Contracting again with $\us^\mu$, and recalling that $\bs\cdot\us = 0$, gives $-b_{\rm n}\cdot \us = \bs\cdot n$, and therefore
\begin{equation}
    \bs^\mu = - \,\, \frac{b_{\rm n}^\mu + (b_{\rm n}\cdot\us) \,\us^\mu}{\us\cdot n}.
    \label{eq:bsurf}
\end{equation}
We can express the gas velocity as
\begin{equation}
    u^\mu = \Gamma \left( \us^\mu + \beta \frac{\bs^\mu}{\bs}\right),
    \label{eq:u_surface}
\end{equation}
where $\bs = |\bs^\mu|$, and find the component of the gas velocity along the surface-frame magnetic field,
\begin{equation}
    \Upsilon = \Gamma \beta = \frac{u\cdot\bs}{\bs}.
\end{equation}
We calculate $\Upsilon$ in the first three cells, use the slope limiter to extrapolate it to the surface [Equation~(\ref{eq:slopelim})] where we recover the scalars
\begin{align}
    \beta &= \frac{\Upsilon}{\sqrt{1 + \Upsilon^2}}\,\mathrm{sign}(\Upsilon),\\
    \Gamma &= \frac{1}{\sqrt{1-\beta^2}},
\end{align}
and then use the local values of $\us^\mu$ and $\bs^\mu$ to construct the surface four-velocity via Equation~(\ref{eq:u_surface}). 

Note that Equation~(\ref{eq:u_surface}) is the 4D generalization of the usual expressions for setting the velocity boundary conditions in non-relativistic simulations \citep[e.g.,][]{2009A&A...508.1117Z},
\begin{align}
    \bm{v}_{\rm p} &= \frac{v_{\rm p}}{B_{\rm p}} \bm{B}_{\rm p}, \\
    v^\phi &= \Omegastar + \frac{v_{\rm p}}{B_{\rm p}} B^\phi,
\end{align}
where $\bm{v}_{\rm p}$ and $\bm{B}_{\rm p}$ are the poloidal three-velocity and magnetic field.

\vspace{5mm}

In surface cells using the force-free boundary condition, we set the boundary gas velocity to the four-velocity of the rotating stellar surface $\us^\mu$ projected such that the coordinate-static observer $\hat{\eta}^\mu$ of Equation~(\ref{eq:coordstaticobs}) only measures a gas velocity perpendicular to the magnetic field,
\begin{equation}
    u^\mu = \Gamma \left(\delta^\mu_\nu - \frac{\hat{b}^\mu \hat{b}_\nu}{\hat{b}^2}\right)\, \us^\nu,
\end{equation}
where $\Gamma$ is given by normalization. In other words, the force-free pulsar wind is launched with no parallel gas velocity at the surface, which is consistent with our hybrid MHD--force-free method as described in Section~\ref{sec:velocity_adjustment}.

\end{document}